\documentclass[11pt]{article}

\usepackage[left=2.5cm,top=4cm,right=2.5cm,bottom=3cm]{geometry}
\usepackage{amsmath,amssymb}
\usepackage{placeins}
\usepackage{float}
\usepackage{physics}
\usepackage{subfigure,epsfig,epsf,color}
\usepackage[utf8x]{inputenc}
\usepackage{amssymb,amsmath}
\usepackage{slashed}
\usepackage{graphicx}
\usepackage{graphics}
\usepackage{epstopdf}
\usepackage{subfigure}
\usepackage{array}
\usepackage{multirow}
\usepackage{amsfonts}
\usepackage{pdfcomment}
\usepackage{hyperref}
\usepackage{cite}
\usepackage{pdflscape}
\usepackage{lipsum}

\begin{document}
\title{Constructing Viable Interacting Dark Matter and Dark Energy Models: A Dynamical Systems Approach}
\maketitle
\begin{center}
Ashmita\footnote{p20190008@goa.bits-pilani.ac.in}, Kinjal Banerjee\footnote{kinjalb@gmail.com} and  Prasanta Kumar Das\footnote{pdas@goa.bits-pilani.ac.in}\\
\end{center}
 
\begin{center}
Birla Institute of Technology and Science-Pilani, K. K. Birla Goa campus, NH-17B, Zuarinagar, Goa-403726, India
\end{center}
\vspace*{0.25in}
\begin{abstract}

We study the evolution of  $k=-1$ FLRW cosmological models for two interacting Dark Matter-Dark Energy Models using dynamical system analysis. Since we are interested in late time evolution, the sign of the interaction term is chosen such that it facilitates the transfer of energy from dark matter to dark energy. We also explore the $k=0$ invariant subspace of these models. We find that both these models have sectors which have a stable fixed point where we can recover an accelerating universe with a negative equation of state. This indicates these can be viable models for our universe. We also rule out certain sectors of these models because they do not give the correct late time observational features. We observe that although we start with a dust-like Dark Matter, its effective equation of state evolves due to its interaction with Dark Energy. As a result, the Dark Matter can display features of stiff matter and exotic matter in the course of evolution.

\end{abstract}

\newpage
\section{Introduction}

The Standard Model of Cosmology, the $\Lambda$CDM model, has been tremendously successful in describing our universe \cite{weinberg2008cosmology,carroll1992cosmological}. This model successfully explains the late-time cosmic acceleration which was first observed using Supernovae (SN) type Ia \cite{riess1998observational,Perlmutter_1999}. Predictions of $\Lambda$CDM model like Baryon Acoustic Oscillations (BAO) \cite{Anderson_2012}, gravitational lensing \cite{treu2003wide,kneib2003wide,bacon2000detection} and polarization of Cosmic Microwave Background (CMB) Spectrum \cite{Kovac2002DetectionOP} have been verified through observations. More recent observations like the Dark Energy Survey (DES) \cite{descollaboration2024darkenergysurveycosmology} and BAO measurements from Dark Energy Spectroscopic Instrument (DESI) \cite{desicollaboration2024desi2024iiibaryon} are also consistent with the predictions of $\Lambda$CDM Model. Interpreting observational data using the $\Lambda$CDM shows that the energy density at current times is dominated by Dark Energy (DE). The matter contribution in the energy density is largely due to Dark Matter (DM). (See \cite{2020,desicollaboration2024desi2024vicosmological,Alam_2021} for exact figures for the relative energy densities from various studies). However there exists some problems which this model has not yet been able to overcome. Some recent reviews about these challenges are \cite{Bull_2016,Perivolaropoulos_2022}.

In the $\Lambda$CDM model, there are six parameters which are determined from observations. The observational data is analyzed assuming some parameters to be constant. For example, the universe is taken to be flat and a cosmological constant, with equation of state $=-1$, is supposed to provide the DE needed for late time acceleration. Usually, observational data is analyzed to check its consistency with $\Lambda$CDM. There may be a large number of models which would satisfy the consistency checks but that will not be feasible computationally. We need some procedure to rule out certain models without extensive data analysis. 

Relaxing assumptions of $\Lambda$CDM usually increases the error bars of the predictions and the tensions among different observational channels become more apparent. Several studies have been done, where in spite of relaxing the assumptions of $\Lambda$CDM, observational bounds  with larger error bars are satisfied. A very incomplete list of recent examples of explorations beyond the standard cosmological model are as follows: 
quintessence model for flat and non flat Friedmann-Lemaitre-Robertson-Walker (FLRW) metric \cite{bhattacharya2024cosmologicalconstraintscurvedquintessence, andriot2024exponentialquintessencecurvedsteep}, dynamical equation of state for DE for various types of FLRW models \cite{perez2024updated}, interacting DM-DE models \cite{li2014large}, non flat models with cosmological constant and various equations of state for Cold Dark Matter (CDM) \cite{rezaei2020bayesian}. A thorough exploration of models beyond $\Lambda$CDM is necessary and therefore is an area of active research. 

In this paper we focus on dynamical DE models. Theoretical problems with a non varying cosmological constant had already given rise to models with dynamical DE. However, in
\cite{weinberg1989cosmological,sahni2000case,COPELAND_2006,carroll2001cosmological,peebles2003cosmological,padmanabhan2003cosmological}, it was shown that assuming a dynamical DE reduces the confidence level of flat universe \cite{Handley_2021,dinda2023constraints}. 
Hence we explore both flat and open FLRW models in this paper. Scalar fields are a convenient description of dynamical DE because of their simplicity, and ability to generate non trivial dynamics \cite{tamanini2014dynamics,gosenca2015dynamical}. There is no fundamental theory predicting the structure and composition of DM but phenomenologically it is necessary for the pressure to be close to zero and the speed of sound of adiabatic perturbations to be sufficiently low to allow cluster formation. One way to include both these features in a model is to have an interacting DM-DE model, where the DE is  modelled as a canonical scalar field and the DM as a perfect fluid. In this work, we take the simplest possible form of DM by modelling it as dust. These two components can exchange energy although the total energy budget of the universe has to remain constant. Each specific model will have its own choice of the potential for the scalar field and a particular form of DM-DE interaction \cite{Arapo_lu_2019,kritpetch2024interacting,Amendola_2001,roy2023exploring,Gumjudpai_2005,boehmer2009dynamics,zhang2011interactions, Amendola_1999,Amendola_2000,Billyard_2000,Holden_2000,Gonzalez_2006,B_hmer_2008,PhysRevD.89.103540,Singh_2016,Bernardi_2017,BISABR2023169443, Mimoso_2006,PhysRevD.99.123520, Chen_2009}. Obviously, any interaction in the DM-DE sector will leave its imprints in the CMB spectrum and in the the large scale structure. However those are more intensive calculations which should be attempted only if a particular DM-DE model reproduces the broad features of our observed cosmological evolution.  
 
Dynamical systems analysis \cite{wainwright1997dynamical,Coley:2003mj,tamanini2014dynamical,Bahamonde_2018,B_hmer_2016} provide a useful tool in this. It provides a quick tool to understand whether key observed features like late-time acceleration are predicted by a model without solving complicated equations. In this method, the equations of motion of a physical system are written as a set of first-order coupled differential equations such that the phase space of the dynamical variables are finite. This allows us to identify key features of the dynamics like fixed points, attractors and stability properties irrespective of initial conditions. Obviously the exact values will be different for different sets of initial conditions but the key features will remain the same. For us this might mean that there exists a stable fixed point of our dynamics which will represent the late time universe. Any model which does not have a late time acceleration in the stable fixed point can then be easily ruled out. Again let us emphasize, having a stable fixed point with such a feature does not necessarily mean that the model is correct but it shows that this model needs to be studied more carefully and passed through tests with observational data. 

In this paper we study a couple of such interacting DM-DE models. The form of the scalar field potential is well studied in literature in the context of inflation and as dynamical DE models. The two forms of DM-DE interactions have also been studied in some contexts but these models have not been analyzed for open universe before. We first study the model for open universe. As a result of having both spatial curvature and DM-DE interactions, our phase space structure is much richer. We also present the results for flat case which is an invariant subspace of our model. An interesting feature of our model is that the interaction with the DE changes the effective equation of state of DM such that it can behave like dust, radiation, stiff matter and even exotic dark matter like quintessence at various stages of the evolution of the universe. This is novel in the sense that it shows even a simple dust like DM can exhibit behaviour of more complicated DM models due to interaction with DE. This feature persists even when we look at the flat universe case. We also present an interesting graphical method to obtain the properties of fixed points which have complicated expressions for the eigenvalues. 

The paper is organized as follows: In Section \ref{section2}, we set up our dynamical system by choosing a set of convenient variables. In section \ref{section3}, we present the analysis of two models having the same scalar potential but two different forms of DM-DE interaction. We end with a discussion of our results in Section \ref{section4}. 

\section{Dynamics of Interacting DM-DE Model} \label{section2}

Homogeneous isotropic solutions of Einstein's equations are given by the FLRW metric \cite{Weinberg:1972kfs, Wald:1984rg}:
 \begin{equation*}
 ds^2=dt^2 - a^2(t)\left[\frac{dr^2}{1-k r^2}+r^2(d\theta^2+\sin^2\theta d\phi^2)\right] 
 \end{equation*}  
where $a(t)$ is the scale factor, $k=0$ for flat models, $k=1$ for closed and $k=-1$ for open universe. In this paper, we will largely concentrate on universes with negative spatial curvature. Since we want to model the late time evolution of the universe, the matter content is described in terms of perfect fluids. Our system will have three types of fluids, one for radiation, one for matter (including Dark Matter) and a minimally coupled scalar field to describe Dark Energy. We also treat curvature as a fluid component with the equation of state $-\frac{1}{3}$ as is conventional in the literature. Although the Standard Model tells that DM and DE are non-interacting, it is worthwhile to explore how the interaction between DM and DE governs the dynamics of the universe. In this work, we analyze situations wherever there is energy exchange between DM and DE. Hence although the total Energy-Momentum tensor is covariantly conserved, energy can flow from the matter to the scalar field. 
The continuity equations for matter($m$), radiation($r$) and scalar field($\phi$) are therefore given by:
 \begin{eqnarray}
    \Dot{\rho}_m + 3 H \rho_m =& Q_I \label{rhom} \\
    \Dot{\rho}_\phi + 3 H \rho_{\phi} (1+\omega_{\phi}) =& -Q_I\label{rhophi}\\
    \Dot{\rho}_r + 3 H \rho_r =& 0 \label{rhor}
\end{eqnarray}
where $\rho_m$, $\rho_r$ and $\rho_\phi$ are the energy densities of matter, radiation and DE scalar field respectively. Here ``dot" denotes derivative with respect to cosmic time $t$, $H = \dot{a}/a$ is the Hubble parameter, $\omega_\phi$ is the equation of state parameter for the scalar field which will depend on the potential energy term of the scalar field Lagrangian. 
Here $Q_I$ (in Eq. (\ref{rhom}) and Eq. (\ref{rhophi})) correspond to DM-DE interactions. To model a dynamical DM-DE interaction where energy flows from the DM to DE, we will need to take $Q_I<0$. Note that $\rho_m$ contains both normal baryonic matter as well as DM. We assume dust like Dark Matter unlike \cite{Deogharia_2021,Pal_2019} so that the entire matter sector EOS is $\omega_m =0$. Since the DM component will dominate, we assume Eq. (\ref{rhom}) and Eq. (\ref{rhophi}) to model energy flowing from DM to DE. Due to the presence  of the $Q_I$ term in Eq. (\ref{rhom}), the standard scaling behaviour of $\rho_m$ with $a$ will no longer hold and the equation of state of DM will change with time.

The exact initial conditions of the universe are not known. Hence the first test of any model is to find the generic features which will be true for all types of initial condition. Dynamical systems \cite{strogatz2018nonlinear} is a mathematical tool which is very useful for that. 
For analyzing a system via dynamical systems, we first need to write down the evolution equation of the degrees of freedom as first-order linear or coupled differential equations. The number of differential equations should be same as the number of degrees of freedom. To analyze the entire phase space, we should try to ensure that none of the degrees of freedom evolve to infinity. In practice, it means we need to take suitable combinations of our physical variables such that their evolutions are first order differential equations and they satisfy some constraint which is valid for all times. The features of a dynamical system which we will be interested in, are the existence of fixed points and their stability. 
The fixed points are those where the first derivatives of all the dynamical variables vanish. The stability properties of each fixed point are determined by the eigenvalues of Jacobian of the system. (See \cite{tamanini2014dynamical,Bahamonde_2018} for reviews on basics of dynamical systems and their applications in cosmology).

The dynamics will depend on the form of $\omega_\phi$ (and hence $V(\phi)$ chosen) as well as the form assumed for $Q_I$. Several possible forms of $V(\phi)$ and $Q_I$ have been explored in \cite{Arapo_lu_2019,kritpetch2024interacting,Amendola_2001,roy2023exploring,Gumjudpai_2005,boehmer2009dynamics,zhang2011interactions, Amendola_1999,Amendola_2000,Billyard_2000,Holden_2000,Gonzalez_2006,B_hmer_2008,PhysRevD.89.103540,Singh_2016,Bernardi_2017,BISABR2023169443, Mimoso_2006,PhysRevD.99.123520,Chen_2009} among others particularly for flat FLRW models. We will compare and contrast the results for our choices of $V(\phi)$ and $Q_I$  with some of these references in the subsequent sections.

First, let us set up our dynamical system. The Einstein equations are 
\begin{eqnarray}
   3 H^2 &= \kappa^2 \bigl( \rho_m + \rho_r + \frac{1}{2}\dot{\phi}^2 +V(\phi) \bigr) - \frac{k}{a^2} 
   \label{friedmanneq} \\
    \Dot{H} &= -\frac{\kappa^2}{2} \biggl[ \rho_m +\frac{4}{3} \rho_r + \Dot{\phi}^2 \biggr] + \frac{k}{a^2}
    \label{accneq}
\end{eqnarray}
where $\kappa^2 = 8\pi G$. 
In this interacting DM and DE setup, the equation of motion for the scalar field is given by
\begin{equation}
    \Ddot{\phi} +3 H \dot{\phi} + \frac{dV}{d\phi} = -\frac{Q_I}{\dot{\phi}}
\end{equation}
A suitable choice of dimensionless variables to construct an autonomous system of equations are:
\begin{equation}
    x = \frac{\kappa \Dot{\phi}}{\sqrt{6} H}, ~ y = \frac{\kappa \sqrt{V}}{\sqrt{3} H}, ~ z = \frac{\sqrt{-k}}{a H}, ~ u = \frac{\kappa \sqrt{\rho_r}}{\sqrt{3} H}, ~ \lambda = -\frac{1}{\kappa } \frac{V_{,\phi}}{V} \label{variables}
\end{equation}
where $V_{,\phi} = \frac{dV(\phi)}{d\phi}$. These variables \footnote{Note that this is not a good set of dynamical variables to study closed universe models because $z$ will become imaginary. So, closed universes have to be studied separately.} have been used in variety of contexts in cosmology in literature \cite{Arapo_lu_2019,kritpetch2024interacting,Amendola_2001,roy2023exploring,Gumjudpai_2005,boehmer2009dynamics,zhang2011interactions,Gonzalez_2006,Amendola_1999,Amendola_2000,Billyard_2000,Holden_2000,B_hmer_2008,PhysRevD.89.103540,Singh_2016,Bernardi_2017,Mimoso_2006,PhysRevD.99.123520,Chen_2009,BISABR2023169443, PhysRevD.60.043501,WEI2012430,LAZAROIU2022115940, Xi-ming_Chen_2009, LAZKOZ2006303, PhysRevD.69.123502,hrycyna2013dynamical,CHATZARAKIS2020168216,PhysRevD.87.084031,Duchaniya_2023,Chatterjee_2024,Kadam_2023,Chatterjee2024,Shah_2019,Frusciante_2014}. 
As is conventional in cosmology, we will change our clock variable from $t$ to $\eta = \ln{a}$. Hence, $\frac{d}{d\eta}=\frac{1}{H}\frac{d}{dt}$. We will use ``prime" to denote derivative with respect to $\eta$. All our evolution plots in the next section will be with respect to $\eta$.

\noindent The autonomous set of differential equations for the dynamical variables is given by 

\begin{subequations} \label{evolutioneqns}
    \begin{flalign}
    x' &= -\frac{3}{2}x + \sqrt{\frac{3}{2}} \lambda y^2 + \frac{3 x}{2} \biggl(  1 + x^2 - y^2 + \frac{u^2}{3} -\frac{z^2}{3} \biggr) -\frac{\kappa Q_I }{\sqrt{6} \dot{\phi} H^2} \label{eq:8a} \\
    y' &= -\sqrt{ \frac{3}{2}} \lambda x y + \frac{3 y}{2} \biggl(  1 + x^2 - y^2 + \frac{u^2}{3} -\frac{z^2}{3} \biggr) \label{eq:8b} \\
    z' &= - z\Biggl[1-\frac{3}{2} \left(  1 + x^2 - y^2 + \frac{u^2}{3} -\frac{z^2}{3} \right)\Biggr] \label{eq:8c} \\
    u' &= -2 u + \frac{3 u}{2} \biggl(  1 + x^2 - y^2 + \frac{u^2}{3} -\frac{z^2}{3} \biggr) \label{eq:8d} \\
    \lambda' &= -\sqrt{6} x \lambda^2 \bigl( \Gamma -1 \bigr); ~ ~ \Gamma = \frac{V V_{,\phi \phi}}{V_{,\phi}^2}  \label{eq:8e}
    \end{flalign}
\end{subequations}
subject to the constraint 
\begin{equation}
    \frac{\kappa^2 \rho_m}{3 H^2} = 1 - (u^2 + x^2 + y^2 + z^2)  
    \label{constraint}
\end{equation}
which is basically the energy balance equation $\Omega_m + \Omega_r + \Omega_\phi + \Omega_k = 1$. Here $\Omega_i$ (with $i = m.~r,~\phi, k $) are the relative energy densities of various degrees of freedom.

The last term of the Eq. \eqref{eq:8a} can be written in terms of the dynamical variables for our specific choices of $Q_I$ in the next sections. To interpret our models, we also need to write the cosmological parameters of interest in terms of the dynamical variables, so that the evolution of those can be studied. They are
\begin{itemize}
 \item Relative energy densities ($\Omega_i$)
 \begin{equation}
    \Omega_{m} = \frac{\kappa^2 \rho_m}{3 H^2} = 1-x^2-y^2-u^2-z^2, ~~ \Omega_{\phi} =  \frac{\kappa^2 \rho_\phi}{3 H^2} =x^2+y^2, ~~ \Omega_r = \frac{\kappa^2 \rho_r}{3 H^2} = u^2, ~~ \Omega_k = \frac{- k}{a^2 H^2} = z^2  \label{relenden}
\end{equation}
 \item Effective equation of state parameter (EOS)
 \begin{equation}
    \omega_{eff} = \frac{p_{eff}}{\rho_{eff}} = \sum_{i} \omega_i \Omega_i = x^2 -y^2 +\frac{u^2}{3} - \frac{z^2}{3} \label{effeqs}
 \end{equation}
 \item Deceleration parameter ($q$) 
 \begin{equation}
    q = -\frac{1}{a}\frac{\Ddot{a}}{\Dot{a}^2} = -1+\frac{3}{2} \biggl(  1 + x^2 - y^2 + \frac{u^2}{3} -\frac{z^2}{3} \biggr) \label{decpar}
\end{equation}
\end{itemize}
In the next section we will look at two explicit models given by two choices of $Q_I$ for a standard choice of the scalar  potential.

\section{Our Models} \label{section3}

To study explicit models, we need to make a choice for the potential of the scalar field $V(\phi)$ and for the interaction term between the DM and DE. We chose an exponential form of the potential is
\begin{equation}
   V = V_0 e^{-\lambda \phi} 
\end{equation}
where $\lambda$ is a positive constant. As we will see later, the range of values for $\lambda$ will depend on what we demand on the existence and the nature of certain fixed points. Note that, our choice of $\lambda$ ranges will depend on the particular evolution history of the universe. Other evolution histories can be constructed by imposing different restrictions on the fixed points and their eigenvalues. 

The choice of the exponential form of the potential is motivated by simplicity. This choice sets $\Gamma=1$ which reduces the number of equations we need to solve from five to four given by Eq. (\ref{evolutioneqns}). Our physical phase space is therefore described by the set: 
\begin{equation}
    \Psi = \{ (x,y,z,u): 0 \le x^2 + y^2 + z^2 + u^2 \le 1, |x| \le 1, y \ge 0, z \ge 0, u \ge 0 \}
\end{equation}

Additionally, the exponential potential allows the existence of scaling solutions where the energy density of the scalar field closely follows that of a perfect fluid\cite{Zhang_2014}. This ensures that neither component becomes negligible at late times. This exponential potential is also one of the simplest example of quintessence and can be supported by theories of gravity such as scalar-tensor theories or string theory\cite{gallego2024anisotropic,amendola1999scaling,holden2000self,alestas2024curve}. Exponential potential is also a strong component of DM in spiral galaxies\cite{guzman1998scalar} and is consistent with observations\cite{huterer1999prospects} of late time acceleration of the universe. 

Having made a choice of $V(\phi)$, we now need to choose a form for $Q_I$. There is no fundamental theory that provides a specific coupling in the dark sector. As a result, any coupling model in this has to be phenomenological. Several forms of $Q_I$ has been studied in literature\cite{Arapo_lu_2019,kritpetch2024interacting,Amendola_2001,roy2023exploring,Gumjudpai_2005,boehmer2009dynamics,zhang2011interactions,Amendola_1999,Amendola_2000,Billyard_2000,Holden_2000,Gonzalez_2006,B_hmer_2008,PhysRevD.89.103540,Singh_2016,Bernardi_2017,BISABR2023169443,Mimoso_2006,PhysRevD.99.123520,Chen_2009}. We will study two such forms:
\begin{enumerate}
 \item $Q_I = \sqrt{\frac{2}{3}} \kappa \beta \dot{\phi} \rho_m $
    \item $Q_I = \beta H \dot{\phi}^2$
\end{enumerate}
The dimensionless parameter $\beta$ which quantifies the strength of the interaction of DM and DE, will depend on what we demand on the existence of certain fixed points. As mentioned above, we will demand $\beta < 0$ because we want energy to flow from DM to DE. As can be seen from the form of the first interaction, the sign of $Q_I$ will also depend on the sign of $\dot{\phi}$ and hence there may be domains of evolution where the energy flows from DE to DM. However at late times it is expected that  $\dot{\phi}$ is positive and then the energy transfer is from DM to DE\footnote{For more realistic models both signs of $\beta$ should be allowed \cite{Pereira_2009}.}. For the second interaction form, energy always flows from DM to DE.

To summarize, our model has two free parameters $\lambda > 0$ and $\beta < 0$. As we shall see below, range of these parameters will be constrained more strongly in each model depending on our choice of the evolution history of the universe. 

\subsection{Model 1: $Q_I = \sqrt{\frac{2}{3}} \kappa \beta \Dot{\phi} \rho_m$}

A coupling of this form can be motivated from scalar tensor theories. It has been studied for various cosmological models for different potentials in literature \cite{Amendola_1999,Amendola_2000,Billyard_2000,Holden_2000,Gonzalez_2006,B_hmer_2008,PhysRevD.89.103540,Singh_2016,Bernardi_2017,BISABR2023169443,doi:10.1142/S0218271815500686}, mostly for flat FRW models to study interacting DM-DE models and also in the context of inflation.

\noindent
Although, we start with a dust-like DM, the interaction with DE changes the EOS of DM. 
The continuity equation for DM given by Eq. ({\ref{rhom}}) can now be written as
\begin{eqnarray}
&&\Dot{\rho}_m + 3 H \rho_m  \left( 1 -\frac{1}{3H} \sqrt{\frac{2}{3}} \kappa \beta \Dot{\phi}\right) = 0
    \nonumber \\
  \mbox{or,} &&
\Dot{\rho}_m + 3 H \rho_m  \left( 1 + \omega_{m_1} \right) = 0 \label{mattereos2} \\
  \Rightarrow && \omega_{m_1} = -\frac{2}{3} \beta x \label{omegam1}
\end{eqnarray}
where $\omega_{m_1}$ is the effective EOS of DM. As we can see, this can be positive, negative or zero depending on the value of $x$ for a given $\eta$. Hence it can mimic, stiff matter, exotic dark matter like k-essence or tachyons, and ordinary dust respectively. 
This leads to several interesting features as we shall see subsequently. 

With this form of $Q_I$, the last term of  Eq. \eqref{eq:8a} becomes $\beta(1-x^2-y^2-z^2-u^2)$.
The fixed points and their stability are summarized in Table (\ref{table:1}). The explicit forms of the eigenvalues are given in Table (\ref{table:9}) in Appendix (\ref{appendix1}). \\

\noindent
As can be seen from Table (\ref{table:1}) below, there are 10 fixed points. The physical description of these fixed points are given in Table (\ref{table:2}).
Let us discuss the properties and stability of each of these fixed points,
\begin{enumerate}
    \item Kinetic domination ($P_k^\pm$): This is a saddle or an {\it unstable} fixed point for a specific ranges of $\beta$ and $\lambda$. If we demand that this is an unstable fixed point, we have to restrict to $\lambda < \sqrt{6}$ and $\beta > -\frac{3}{2}$. We will restrict ourselves to this range of parameters in the rest of the discussion. In that case we can say the universe evolves due to the kinetic term of the scalar field as can be seen from the $x$ value corresponding to that point. (See Equation (\ref{variables})). This observation indicates that that the universe started from an early kination epoch.

    \item Radiation domination ($P_r$): This is a saddle point where $\omega_{eff} =\frac{1}{3}$ and $\Omega_r =1$. This indicates that it is a radiation dominated phase. 

    \item Curvature domination ($P_c$): This is again a saddle point where $\Omega_k =1$, indicating it is a curvature dominated regime.

\begin{table}[H]
\addtolength{\tabcolsep}{-4pt}
\renewcommand{\arraystretch}{2.2}
\centering
\begin{tabular}{| c | c | c | c | c | c | c | c | c | c | c | c |} 
 \hline
Point & x & y & z & u & Existence & Stability \\ [0.5ex] 
\hline\hline
& & & & & & $P^-_k$: Fully unstable \\

$P^\pm_k$ & $\pm 1$ & $0$ & $0$ & $0$ & $\forall ~ \lambda$ & $P^+_k$: When $\beta \le -\frac{3}{2}$,  Saddle $\forall ~ \lambda$ \\ [0.5ex]
& & & & & &$\beta > -\frac{3}{2}$ \multirow{2}{*}{\(\left\{\begin{array}{c} \\ \\ \end{array}\right.\)} Saddle point for $\lambda \ge \sqrt{6}$  \\
& & & & & & ~ ~ ~ ~ ~ ~ ~ Fully unstable for $\lambda<\sqrt{6}$  \\
\hline
$P_r$ & $0$ & $0$ & $0$ & $ 1$ & $\forall ~ \lambda$ & Saddle \\ [0.5ex]
\hline
$P_c$ & $0$ & $0$ & $1$ & $0$ & $\forall ~ \lambda$ & Saddle \\ [0.5ex]
\hline
$P_{r \phi}$ & $\frac{1}{\lambda} \sqrt{\frac{8}{3}}$ & $\frac{2}{\sqrt{3} \lambda}$ & $0$ & $\sqrt{1-\frac{4}{\lambda^2}}$ & $\lambda>2$ & Unstable Spiral \\ [0.5ex]
\hline
$P_{\phi}$ & $\frac{\lambda}{\sqrt{6}}$ & $\sqrt{1-\frac{\lambda^2}{6}}$ & $0$ & $0$ & $\lambda<\sqrt{6}$ & Stable for $\lambda \le \sqrt{2}$ \\ [0.5ex]

& & & & & & Saddle for $\lambda > \sqrt{2}$ \\
\hline
$P_{{mixed}_1}$ & $-\frac{1}{2\beta}$ & $0$ & $0$ & $\sqrt{1-\frac{3}{4 \beta^2}}$ & $\forall  \lambda$, $\beta^2 > \frac{3}{4}$ & Saddle \\ [0.5ex]
\hline
$P_{{mixed}_2}$ & $\frac{1}{2\beta}$ & $0$ & $\sqrt{1+\frac{3}{4 \beta^2}}$ & $0$ & $\forall ~ \lambda$ & Saddle  \\ [0.5ex]
\hline
$P_{{m \phi}_1}$ & $-\frac{2 \beta}{3}$ & $0$ & $0$ & $0$ & $\forall ~ \lambda$ & Saddle \\ [0.5ex]
\hline
$P_{{m \phi}_2}$ & $\frac{3}{\sqrt{6} \lambda +2 \beta}$ & $ \frac{\sqrt{4\beta^2+2\sqrt{6}\lambda \beta + 9}}{\sqrt{6} \lambda +2 \beta}$ & $0$ & $0$ & $\forall ~ \lambda$ & Saddle \\ [0.5ex]
\hline

$P_{c \phi}$ & $\frac{1}{\lambda} \sqrt{\frac{2}{3}}$ & $\frac{2}{\sqrt{3} \lambda}$ & $\sqrt{1-\frac{2}{\lambda^2}}$ & $0$ & $\lambda>\sqrt{2}$ & Stable Spiral \\ [0.5ex]

\hline
\end{tabular}
\caption{Fixed points for Model $1$  and their stability in the parameter space of $\lambda$ and $\beta$ for $k=-1$.}
\label{table:1}
\end{table}

    \item Radiation scaling ($P_{r\phi}$): This is an {\it unstable} spiral. The energy density gets contribution from both the scalar field as well as radiation. However, $\omega_{eff}=\frac{1}{3}$ which mimics a pure radiation dominated universe. Note that, when $\lambda=2$, this point merges with the $P_\phi$.

    \item Scalar field domination ($P_\phi$): This can either be a saddle or a {\it stable} point, depending on the range of $\lambda$. 
    Here, the energy density is dominated by the scalar field since $\Omega_\phi =1$. If we consider the range of $\lambda$ for which it is a stable fixed point then $\omega_{eff}$ is negative, while for the other range it can be both positive or negative.
    
   \item Mixed phase ($P_{{mixed}_1}$): This is a saddle point which will exist only if $\beta^2 > \frac{3}{4}$. This fixed point is a new feature of our model and exists only because of the specific form of the interaction chosen. Although the energy contribution comes from scalar field, matter and radiation, $\omega_{eff}=\frac{1}{3}$ which shows that it mimics the radiation domination. If we take $\beta^2 = \frac{3}{4}$, this point merges with the point $P_{{m\phi}_1}$.
      
    \item Matter scaling ($P_{{m \phi}_1}$ ): This is a saddle point which exist due to consideration of a particular interaction and our potential. The evolution of the universe is driven by both the scalar field and matter. 

    \item ($P_{{mixed}_2}$ and $P_{{m \phi}_2}$): Again, these two saddle points exist due to the form of interaction and the potential choice. These points give negative relative energy density for matter. In case of $P_{{mixed}_2}$, the evolution of universe is driven due to matter, scalar field and curvature while for $P_{{m \phi}_2}$ the evolution of universe is driven by just matter and scalar field. Explicit analysis of $P_{{m \phi}_2}$ is presented in Appendix (\ref{appendix1}). 

    \item Curvature scaling ($P_{c\phi}$): This is a {\it stable} spiral where the energy density gets contribution from both curvature and the scalar field. The corresponding $\omega_{eff}=-\frac{1}{3}$ which mimics the pure curvature domination epoch of the universe. For $\lambda=\sqrt{2}$, this merges with the point $P_\phi$.
\end{enumerate}

Note that the stability analysis reported in Table (\ref{table:1}) is with the choice of $- \frac{3}{2} < \beta < 0$ because we want the point $P_{k}^{\pm}$ to be an unstable fixed point. \\

\begin{figure}[H]
\centering
    \begin{minipage}[t]{0.44\textwidth}
        \centering
        \includegraphics[width=\textwidth]{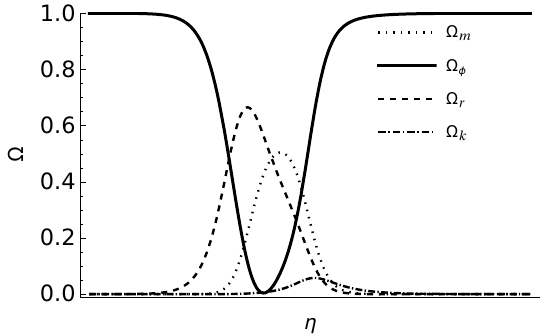} 
    \end{minipage}
    \hspace{0.5cm}
    \begin{minipage}[t]{0.44\textwidth}
        \centering
        \includegraphics[width=\textwidth]{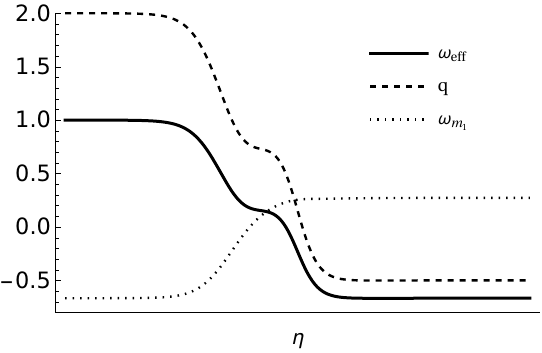}
    \end{minipage}
    
    \vspace{0.5cm}
    
    \begin{minipage}[t]{0.44\textwidth}
        \centering
        \includegraphics[width=\textwidth]{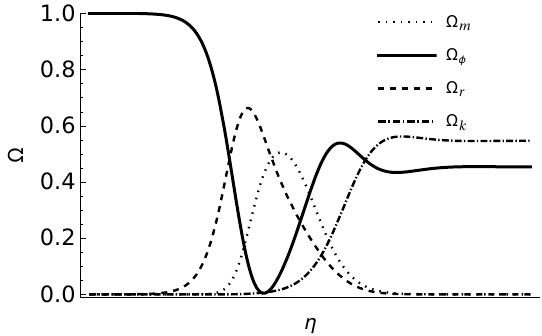} 
    \end{minipage}
    \hspace{0.5cm}
    \begin{minipage}[t]{0.44\textwidth}
        \centering
        \includegraphics[width=\textwidth]{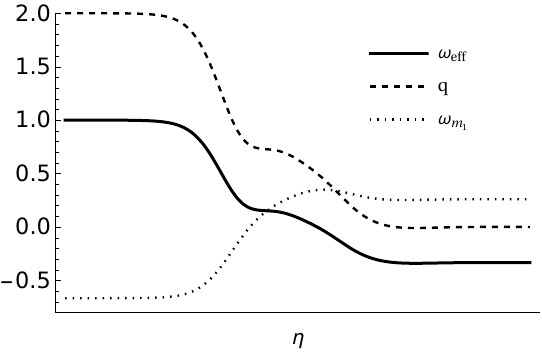} 
    \end{minipage}
    \caption{Evolution plot for the different cosmological parameters. Here, the first row represents the plots for $\lambda=1$ and $\beta=-1$ while the second row represents the plots for $\lambda=2$ and $\beta=-1$.}
    \label{figuree1}
\end{figure}

From Table (\ref{table:1}), we have two possible ranges of $\lambda$ for the evolution of universe. 
\begin{itemize}
\item  {\bf{Scenario 1}}: If $\lambda \le \sqrt{2}$, the universe starts with a kination epoch $P_k^\pm$ and reaches field domination epoch $P_{\phi}$ which is a stable point with all other epochs like $P_r$, $P_{r\phi}$, $P_{{m \phi}_1}$, $P_{{m \phi}_2}$, $P_{{mixed}_1}$, $P_{{mixed}_2}$, $P_{c\phi}$, $P_c$ as saddle points.  
The evolution of the cosmological parameters are plotted in the first row of Fig. (\ref{figuree1}). It is plotted for a generic value of the model parameters (here we have taken $\beta=-1$ and $\lambda=1$). This plot shows that the universe starts evolving due to field domination showed by thick black line, then radiation domination occurs which is shown by black dashed line and after that there is matter domination which is shown by black dotted line. At late times, field starts dominating. Note that the contribution arising from curvature which is shown by black dashed-dotted line is negligible in this case. We see that EOS and $q$ are negative for late times thereby making this a viable model of our universe. 
The behaviour of $\omega_{m_1}$ is plotted in the first row of Fig. (\ref{figuree1}). It acts like exotic dark matter during early epochs, dust in the intermediate times and like radiation at late times.

\item {\bf {Scenario 2}}: Another possibility is if $\sqrt{2}<\lambda<\sqrt{6}$, universe starts with the kination epoch and reaches $P_{c\phi}$ with all other epochs like $P_r$, $P_{r\phi}$, $P_{{m \phi}_1}$, $P_{{m \phi}_2}$,  $P_{{mixed}_1}$, $P_{{mixed}_2}$, $P_{\phi}$, $P_c$ as saddle points. In the second row of Fig. (\ref{figuree1}), we again have plotted the evolutions of different energy densities, EOS and $q$ for the same $\beta(=-1)$ but $\lambda$ is chosen from this range ($\lambda(=2)$).  
We see that now the universe starts from field domination then there is radiation domination, matter domination and again field domination but after that there is contribution from curvature too. EOS and $q$ for the same $\lambda$ and $\beta$ are plotted. We can see that although EOS is negative at late times but $q =0$. So this scenario is not consistent with late time acceleration which we observe in our universe. But the behaviour of $\omega_{m_1}$ is similar to the one we have in previous scenario.
\end{itemize}

\noindent
A recent work \cite{andriot2024exponentialquintessencecurvedsteep} for open universe has studied the same $V(\phi)$ but with $Q_I=0$. The addition of the interaction term gives rise to several saddle points $P_{{m \phi}_1}$, $P_{{m \phi}_2}$, $P_{{mixed}_1}$ and $P_{{mixed}_2}$. The matter domination phase in \cite{andriot2024exponentialquintessencecurvedsteep} missing from our model can be recovered from $P_{{m \phi}_1}$, $P_{{m \phi}_2}$ by setting the interaction parameter $\beta = 0$, while the $P_{{mixed}_1}$ and $P_{{mixed}_2}$ points would not exist anymore. Hence, our model, while being consistent with previous results, have a much richer phase space structure. As can be seen from Table (\ref{table:9}) of the Appendix (\ref{appendix1}), the stability conditions depend on both $\beta$ and $\lambda$ but they reduce to the the cases given in \cite{andriot2024exponentialquintessencecurvedsteep} when $\beta = 0$. 
\begin{table}[H]
\addtolength{\tabcolsep}{-0.8pt}
\renewcommand{\arraystretch}{2.2}
\centering
\begin{tabular}{| c | c | c | c | c | c | c | c | c | c | c | c |} 
 \hline
Point & $\Omega_m$ & $\Omega_r$ & $\Omega_\phi$ & $\Omega_k$ & 
$\omega_{eff}$  & q & $\omega_{m_1}$\\
\hline\hline

$P^\pm_k$ & $0$ & $0$ & $1$ & $0$ & $1$ & $2$ & $\mp \frac{2\beta}{3}$\\ [0.5ex]
\hline
$P_r$ & $0$ & $1$ & $0$ & $0$ & $\frac{1}{3}$ & $1$ & $0$ \\ [0.5ex]
\hline
$P_c$ & $0$ & $0$ & $0$ & $1$ & $-\frac{1}{3}$ & $0$ & $0$ \\ [0.5ex]
\hline
$P_{r \phi}$ & $0$ & $1-\frac{4}{\lambda^2}$ & $\frac{4}{\lambda^2}$ & $0$ & $\frac{1}{3}$ & $1$ & $-\frac{4\beta}{3\lambda}\sqrt{\frac{2}{3}}$ \\ [0.5ex]
\hline
$P_{\phi}$ & $0$ & $0$ & $1$ & $0$ & $-1+\frac{\lambda^2}{3}$ & $-1+\frac{\lambda^2}{2}$ & $-\frac{1}{3}\sqrt{\frac{2}{3}}\lambda\beta$\\ [0.5ex]
\hline
$P_{{mixed}_1}$ & $\frac{1}{2\beta^2}$ & $1-\frac{3}{4 \beta^2}$ & $\frac{1}{4\beta^2}$ & $0$ & $\frac{1}{3}$ & $1$ & $\frac{1}{3}$ \\ [0.5ex]
\hline
$P_{{mixed}_2}$ & $-\frac{1}{\beta^2}$ & $0$ & $\frac{1}{4\beta^2}$ & $1+\frac{3}{4 \beta^2}$ & $-\frac{1}{3}$ & $0$ & $-\frac{1}{3}$ \\ [0.5ex]
\hline
$P_{{m \phi}_1}$ & $1-\frac{4\beta^2}{9}$ & $0$ & $\frac{4\beta^2}{9}$ & $0$ & $\frac{4\beta^2}{9}$ & $\frac{1}{6}(3+4 \beta^2)$ & $\frac{4\beta^2}{9}$ \\ [0.5ex]
\hline
$P_{{m \phi}_2}$ & $\frac{2 \left(3 \lambda^2+\sqrt{6} \beta \lambda-9\right)}{\left(2 \beta +\sqrt{6} \lambda\right)^2}$ & $0$ & $\frac{2 \left(2 \beta^2+\sqrt{6} \beta  \lambda+9\right)}{\left(2 \beta +\sqrt{6} \lambda\right)^2}$ 
& $0$ & $-\frac{2\beta}{\sqrt{6} \lambda +2 \beta}$ & $\frac{\sqrt{6} \lambda-4\beta}{2\sqrt{6} \lambda +4 \beta}$ & $\frac{-2\beta}{\sqrt{6} \lambda +2 \beta}$ \\ [0.5ex]
\hline
$P_{c \phi}$ & $0$ & $0$ & $\frac{2}{\lambda^2}$ & $1-\frac{2}{\lambda^2}$ & $-\frac{1}{3}$ & $0$ & $-\frac{2}{3} \sqrt{\frac{2}{3}} \frac{\beta}{\lambda}$ \\ [0.5ex]

\hline
\end{tabular}
\caption{List of the cosmological parameters for Model $1$.}
\label{table:2}
\end{table}
\noindent
Let us look at a few interesting sub-case of our model:

\subsubsection{No Radiation}

Another interesting invariant subspace can be constructed when there is no radiation contribution while retaining the curvature component. The fixed points and their behaviour is given in Table (\ref{table:3}). In doing so, we can see few fixed points $P_r$, $P_{r\phi}$, $P_{{mixed}_1}$ are no longer there. This is expected because from Table (\ref{table:2}), we can see the $\omega_{eff}$ was behaving like radiation for these points. We have two points which are stable $P_{c\phi}$ and $P_{\phi}$, hence we have two possible histories of evolution from unstable point $P_k^{\pm}$ which exists as long as  $-\frac{3}{2}< \beta < 0$ and $0 < \lambda < \sqrt{6}$.

\begin{table}[H]
\addtolength{\tabcolsep}{-6.0pt}
\renewcommand{\arraystretch}{2.3}
\centering
\begin{tabular}{| c | c | c | c | c | c | c | c | c | c | c | c |} 
 \hline
Point & Stability & $\omega_{eff}$ & $\Omega_m$ & $\Omega_\phi$ & $\Omega_k$ & $q$ & $\omega_{m_1}$\\ [0.5ex] 
\hline\hline

& $P^-_k$: Fully unstable &  &  &  &  &  &\\

&  $P^+_k$: When $\beta \le -\frac{3}{2}$,  Saddle $\forall ~ \lambda$ &  &  &  &  &  & \\ 

$P^\pm_k$ & $\beta > -\frac{3}{2}$ \multirow{2}{*}{\(\left\{\begin{array}{c} \\ \end{array}\right.\)} Saddle point for $\lambda \ge \sqrt{6}$  & $1$ & $0$  & $1$ & $0$ & $2$ & $\mp \frac{2\beta}{3}$ \\ [0.5ex]

& ~ ~ ~ ~ ~ ~ ~ Fully unstable for $\lambda<\sqrt{6}$ & & & & & &\\
\hline

$P_c$ & Saddle & $-\frac{1}{3}$ & $0$ & $0$ & $1$ & $0$ & $0$\\ [0.5ex]
\hline

$P_{c\phi}$ & Stable (Spiral) for $\lambda>\sqrt{2}$ & $-\frac{1}{3}$ & $0$ & $\frac{2}{\lambda^2}$ & $1-\frac{2}{\lambda^2}$ & $0$ & $-\frac{2}{3} \sqrt{\frac{2}{3}} \frac{\beta}{\lambda}$ \\ 
& Otherwise $P_{c\phi}$ won't exist & & & & & &\\
\hline

$P_\phi$ & Stable for $\lambda \le \sqrt{2}$ & $-1+\frac{\lambda^2}{3}$ & $0$ & $1$ & $0$ & $-1+\frac{\lambda^2}{2}$ & $-\frac{\sqrt{2}}{3\sqrt{3}}\lambda\beta$ \\ [0.5ex]
 & Saddle for $\lambda > \sqrt{2}$ &  &  &  &  &  & \\
\hline

$P_{{mixed}_2}$ & Saddle & $-\frac{1}{3}$ & $-\frac{1}{\beta^2}$ & $\frac{1}{4\beta^2}$ & $1+\frac{3}{4 \beta^2}$ & $0$ & $-\frac{1}{3}$ \\ [0.5ex]
\hline

$P_{{m\phi}_1}$ & Saddle & $\frac{1}{6}(3+4\beta^2)$ & $1-\frac{4\beta^2}{9}$ &  $\frac{4\beta^2}{9}$ & $0$ & $\frac{1}{6}(3+4 \beta^2)$ & $\frac{4\beta^2}{9}$\\ [0.5ex]
\hline

$P_{{m\phi}_2}$ & Saddle & $\frac{-2\beta}{\sqrt{6} \lambda +2 \beta}$ & $\frac{2 \left(3 \lambda^2+\sqrt{6} \beta \lambda-9\right)}{\left(\sqrt{6} \lambda +2 \beta\right)^2}$ & $\frac{2 \left(2 \beta^2+\sqrt{6} \beta  \lambda+9\right)}{\left(\sqrt{6} \lambda +2 \beta\right)^2}$ & $0$ & $\frac{\sqrt{6} \lambda-4\beta}{2\sqrt{6} \lambda +4 \beta}$ & $\frac{-2\beta}{\sqrt{6} \lambda +2 \beta}$\\ [0.5ex]

\hline
\end{tabular}
\caption{List of fixed points for Model $1$ and its stability in the parameter space of $\lambda$ and $\beta$ for no radiation.} 
\label{table:3}
\end{table}
\begin{itemize}
    \item If $\lambda \le \sqrt{2}$, the universe starts with a kinetic dominated epoch $P_k^\pm$ and reaches the field domination phase $P_{\phi}$ which is a stable point with all other epochs like  $P_c$, $P_{c\phi}$, $P_{{mixed}_2}$, $P_{{m \phi}_1}$, $P_{{m \phi}_2}$ as saddle points. This scenario shows late time acceleration.
    
    \item If $\sqrt{2}<\lambda<\sqrt{6}$, universe starts with the kinetic domination and reaches $P_{c\phi}$ with all other epochs like $P_c$, $P_{\phi}$, $P_{{mixed}_2}$,  $P_{{m \phi}_1}$, $P_{{m \phi}_2}$ as saddle points. Here, $q=0$ at the stable fixed point indicating that this is not a viable model for late time acceleration of the universe.
\end{itemize}
For the same potential, one recent work\cite{alestas2024curvecurvecurvatureassistedquintessence} is done where they have no radiation component as well along with no interaction. Our model has similar results to their work if we switch off the interaction. 

\subsubsection{Flat universe:~$k = 0$}

The $k=0$ case can be obtained as an invariant subspace of our model by setting $z= 0$. In this case the phase space is given by the constraints ($z=0, x^2+y^2+u^2+\Omega_m=1$). For this case the fixed points and their stability changes. These are tabulated in Table (\ref{table:4}). 

\begin{table}[H]
\addtolength{\tabcolsep}{-5.5pt}
\renewcommand{\arraystretch}{2.3}
\centering
\begin{tabular}{| c | c | c | c | c | c | c | c | c | c | c |} 
 \hline
Point & Stability & $\omega_{eff}$ & $\Omega_m$ & $\Omega_r$ & $\Omega_\phi$ & $q$ & $\omega_{m_1}$ \\ [0.5ex] 
\hline\hline
&  $P^-_k$: Fully unstable &  &  &  & & &  \\

& $P^+_k$: When $\beta \le -\frac{3}{2}$,  Saddle $\forall ~ \lambda$ & & & & & & \\ 

$P^\pm_k$ & $\beta > -\frac{3}{2}$ \multirow{2}{*}{\(\left\{\begin{array}{c} \\ \\ \end{array}\right.\)} Saddle point for $\lambda \ge \sqrt{6}$  & $1$ & $0$ & $0$ & $1$ & $2$ & $\mp \frac{2\beta}{3}$ \\ [0.5ex]

&  ~ ~ ~ ~ ~ ~ ~ Fully unstable for $\lambda<\sqrt{6}$ & & & & & & \\
\hline

$P_r$ & Saddle & $\frac{1}{3}$ & $0$ & $1$ & $0$ & $1$ & $0$\\ [0.5ex]
\hline

$P_{r\phi}$ & Unstable spiral  & $\frac{1}{3}$ & $0$ & $1-\frac{4}{\lambda^2}$ & $\frac{4}{\lambda^2}$ & $1$ &  $-\frac{4\beta}{3\lambda}\sqrt{\frac{2}{3}}$\\ [0.5ex]
\hline

$P_\phi$ & Stable for $\lambda \le 2$ & $-1+\frac{\lambda^2}{3}$ & $0$ & $0$ & $1$ & $-1+\frac{\lambda^2}{2}$ & $-\frac{\sqrt{2}}{3\sqrt{3}}\lambda\beta$ \\ [0.5ex]
&  Saddle for $\lambda > 2$ & & & & & & \\
\hline

$P_{{mixed}_1}$ & Saddle for $\beta^2 
\ge \frac{3}{4}$ & $\frac{1}{3}$ & $\frac{1}{2\beta^2}$ & $1-\frac{3}{4 \beta^2}$ & $\frac{1}{4\beta^2}$ & $1$ & $\frac{1}{3}$ \\ [0.5ex]
\hline

$P_{{m\phi}_1}$ & Saddle & $\frac{4\beta^2}{9}$ & $1-\frac{4\beta^2}{9}$ & $0$ & $\frac{4\beta^2}{9}$ & $\frac{1}{6}(3+4 \beta^2)$ & $\frac{4\beta^2}{9}$ \\ [0.5ex]
\hline

$P_{{m\phi}_2}$ & Saddle & $\frac{-2\beta}{\sqrt{6} \lambda +2 \beta}$ & $\frac{2 \left(3 \lambda^2+\sqrt{6} \beta \lambda-9\right)}{\left(\sqrt{6} \lambda +2 \beta\right)^2}$ & $0$ & $\frac{2 \left(2 \beta^2+\sqrt{6} \beta  \lambda+9\right)}{\left(\sqrt{6} \lambda +2 \beta\right)^2}$ & $\frac{\sqrt{6} \lambda-4\beta}{2\sqrt{6} \lambda +4 \beta}$ & $\frac{-2\beta}{\sqrt{6} \lambda +2 \beta}$ \\ [0.5ex]

\hline
\end{tabular}
\caption{List of fixed points for Model $1$ and their stability in the parameter space of $\lambda$ and $\beta$ for $k = 0$.}
\label{table:4}
\end{table}
As expected the fixed points, where $\omega_{eff} = -\frac{1}{3}$ disappear. So, in terms of the fixed points tabulated in Table (\ref{table:1}),  $P_c$, $P_{c\phi}$, $P_{{mixed}_2}$ are no longer there. In the $k= -1$ case, we had two possible histories of evolution. Here the stable spiral $P_{c\phi}$, disappears, leaving only one stable point $P_{\phi}$. Additionally, the stability conditions for this point have also changed. Now to evolve the universe from the unstable point $P_k^{\pm}$ to the stable point $P_{c\phi}$ via the other saddle points, the range of parameters become $-\frac{3}{2}< \beta < 0$ and $0 < \lambda < 2$. If we want to ensure $q<0$ at late times we need to restrict the range of $\lambda$ further to $0 < \lambda < \sqrt{2}$. These results are similar to \cite{Bernardi_2017} who work with the same form of scalar potential but with a generic interaction term $Q$. 

\begin{figure}[H]
    \begin{minipage}[t]{0.45\textwidth}
        \centering
        \includegraphics[width=\textwidth]{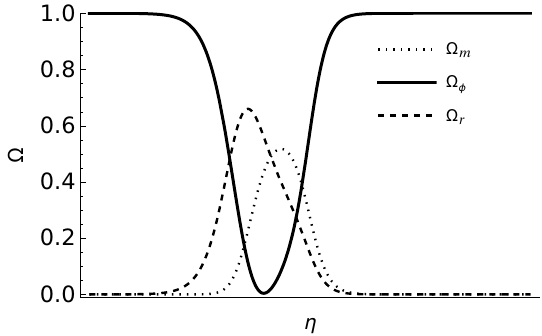} 
    \end{minipage}
    \hspace{0.5cm}
    \begin{minipage}[t]{0.45\textwidth}
        \centering
        \includegraphics[width=\textwidth]{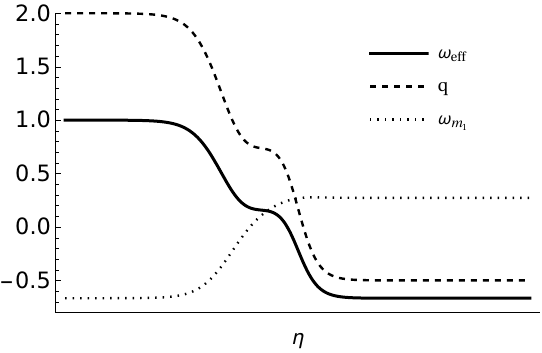}
    \end{minipage}
    \caption{Evolution plot for the different cosmological parameters when $k=0$ for $\lambda=1$ and $\beta=-1$. }
    \label{figureevolz}
\end{figure}
When we switch off the curvature term, we see that there is no peak from curvature in the evolution plot given by Fig. (\ref{figureevolz}). Again we observe that the scalar field drives the evolution, followed by a phase of radiation domination, and subsequently transitioning into matter domination. Then, scalar field begins to dominate, leading to the acceleration at late times.

One interesting feature of this interacting DM-DE model is that it can mimic effects of exotic DM equations of state. Changing EOS of DM has already been observed in \cite{deogharia2021generalized,roy2024dynamicalstabilityphasespace,Hussain_2023}. In those studies, the change was more apparent at late times, while in our model $\omega_{m_1}$ starts of by mimicing an exotic DM at early times and smoothly transitions to behaving like radiation at late times. (See Fig. (\ref{figuree1}) for $k=-1$ and Fig. (\ref{figureevolz}) for $k=0$).

\subsection{Model 2: $Q_I = \beta H \dot{\phi}^2$}
\noindent
This particular form of interaction was first studied in \cite{Mimoso_2006} to analyse the asymptotic behavior of the warm inflation scenario with viscous pressure. In \cite{PhysRevD.99.123520}, the same interaction term was studied, focusing solely on DM and DE for different types of exponential potential for flat universe. 
The continuity equation given by Eq. ({\ref{rhom}}) can now be written as
\begin{eqnarray}
  &&
 \Dot{\rho}_m + 3 H \rho_m \left( 1 - \frac{2\beta}{3}\frac{x^2}{1-x^2-y^2-z^2-u^2}  \right) = 0  \nonumber \\
  \mbox{or,} &&  \Dot{\rho}_m + 3 H \rho_m \left( 1 + \omega_{m_2} \right) = 0 \label{mattereos2}\\
  \Rightarrow && \omega_{m_2} = -\frac{2\beta}{3}\frac{x^2}{1-x^2-y^2-z^2-u^2} \label{omegam2}
\end{eqnarray}

\noindent where $\omega_{m_2}$ is the effective EOS of DM for the given interaction form. As we can see, similar to Model $1$, this can be positive, negative or zero depending on the values of $\beta$ and the dynamical variables for a given $\eta$ and hence can mimic various types of equations of state. As can be seen from Eq. (\ref{omegam2}), $\omega_{m_2}$ blows up when the denominator becomes zero. At these points, the constraint continues to hold but $\omega_{m_2}$ is not a good descriptor of the behaviour of DM. 
 
For this type of interaction the last term of Eq. \eqref{eq:8a} becomes $\beta x$. The list of fixed points and their stability is summarized in Table (\ref{table:5}). The physical parameters corresponding to these fixed points are given in Table (\ref{table:6}). 
\begin{table}[H]
\addtolength{\tabcolsep}{-5.2pt}
\renewcommand{\arraystretch}{2.2}
\centering
\begin{tabular}{| c | c | c | c | c | c | c | c | c | c | c | c |} 
 \hline
Point & x & y & z & u & Existence & Stability \\ [0.5ex] 
\hline\hline
$P_{{m \phi}_1}^\pm$ & $\pm \sqrt{1+\frac{2 \beta}{3}}$ & $0$ & $0$ & $0$ & $\forall ~ \lambda, \beta \ge -\frac{3}{2}$ & $P^+_{{m\phi}_1}$: \multirow{2}{*}{\(\left\{\begin{array}{c} \\ \\ \end{array}\right.\)} Unstable for $\lambda<\frac{6+2\beta}{\sqrt{6+4\beta}}$ \\ [0.5ex]

&  &  &  &  &  & ~ ~ ~ ~ ~ ~ ~ Saddle for $\lambda \ge \frac{6+2\beta}{\sqrt{6+4\beta}}$  \\

&  &  &  &  &  & $P^-_{{m\phi}_1}$: Fully unstable \\
\hline

$P_m$ & $0$ & $0$ & $0$ & $0$ & $\forall ~ \lambda$ & Saddle \\
\hline

$P_r$ & $0$ & $0$ & $0$ & $ 1$ & $\forall ~ \lambda$ & Saddle \\ [0.5ex]
\hline

$P_c$ & $0$ & $0$ & $1$ & $0$ & $\forall ~ \lambda$ & Saddle \\ [0.5ex]
\hline

$P_{mixed_1}$ & $\frac{1}{\lambda} \sqrt{\frac{2}{3}}$ & $\sqrt{\frac{2}{3}} \frac{\sqrt{2+\beta}}{\lambda}$ & $ \sqrt{1-\frac{2(1+\beta)}{\lambda^2}}$ & $0$ & $\lambda > \sqrt{2(1+\beta)}$, & Saddle $(-1,\lambda_2,\lambda_3,\lambda_4)$ \\ 

& & & & & $\beta > -1$ & See Fig. (\ref{figure1}) first row\\
[0.5ex]
\hline

$P_{mixed_2}$ & $\frac{1}{\lambda} \sqrt{\frac{8}{3}}$ & $\frac{ 2 \sqrt{1+\beta}}{\sqrt{3}\lambda}$ & $0$ & $\sqrt{1-\frac{4(1-\beta)}{\lambda^2}}$ & $\lambda > \sqrt{4(1-\beta)},$ & Saddle $(1,\lambda_2,\lambda_3,\lambda_4)$ \\  [0.5ex]

& & & & & $\beta > -1$ & See Fig. (\ref{figure1}) second row \\
\hline

$P_{{m \phi}_2}$ & $P_1$ & $P_2$ & $0$ & $0$ & Refer to Fig. (\ref{figure2}) & Stable \\ [0.5ex]
\hline

$P_{{m \phi}_3}$ & $P_3$ & $P_4$ & $0$ & $0$ & Refer to Fig. (\ref{figure3}) & Saddle \\ [0.5ex]

\hline
\end{tabular}
\caption{List of the fixed points and their stability  in the parameter space of $\lambda$ and $\beta$ for Model $2$ for $k=-1$.} 
\label{table:5}
\end{table}
In Table (\ref{table:5}), $P_1$, $P_2$, $P_3$, $P_4$ correspond to 
\begin{eqnarray}
    &&P_1 = \frac{3+\lambda^2+\beta-\delta}{2\sqrt{6} \lambda} 
    \label{p1} \\
    &&P_2 = \frac{\sqrt{-\lambda^4 +\lambda^2 \left(6+\delta \right)-(3+\beta) \left(-3-\beta+\delta \right)}}{2\sqrt{3} \lambda} 
    \label{p2} \\
    &&P_3 = \frac{3+\lambda^2+\beta+\delta}{2\sqrt{6} \lambda} 
    \label{p3} \\
    &&P_4 = \frac{\sqrt{-\lambda^4 -\lambda^2 \left(-6+\delta \right)-(3+\beta) \left(3+\beta+\delta\right)}}{2\sqrt{3} \lambda}
    \label{p4}
\end{eqnarray}

\noindent
where $ \delta = \sqrt{\lambda^4+ 2\lambda^2 (-3+\beta)+ (3+\beta)^2}$.
As can be seen from the above expression, the domains of $\lambda$  and $\beta$ for which these fixed points exist are very complicated. Even when they exist, the analysis of their stability depending on the nature of the eigenvalues is even more complicated. \\

\noindent Instead of trying to analyze these issues analytically we try to address these issues graphically. 
We choose to ensure that $P_{m\phi_1}^+$ becomes a fully unstable fixed point. This gives the following ranges

\begin{equation}
   -\frac{3}{2} < \beta < 0 ~ ~ ~; ~ ~ ~ 0 < \lambda<\frac{6+2\beta}{\sqrt{6+4\beta}}  \label{rangem2}
\end{equation}

\begin{figure}[H]
\centering
    \begin{minipage}[t]{0.325\textwidth}
        \centering
        \includegraphics[width=\textwidth]{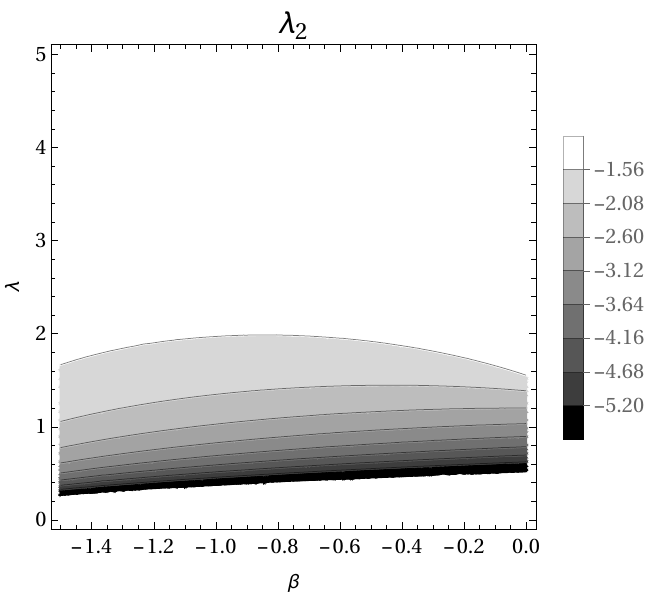}
    \end{minipage}%
    \hfill
    \begin{minipage}[t]{0.325\textwidth}
        \centering
        \includegraphics[width=\textwidth]{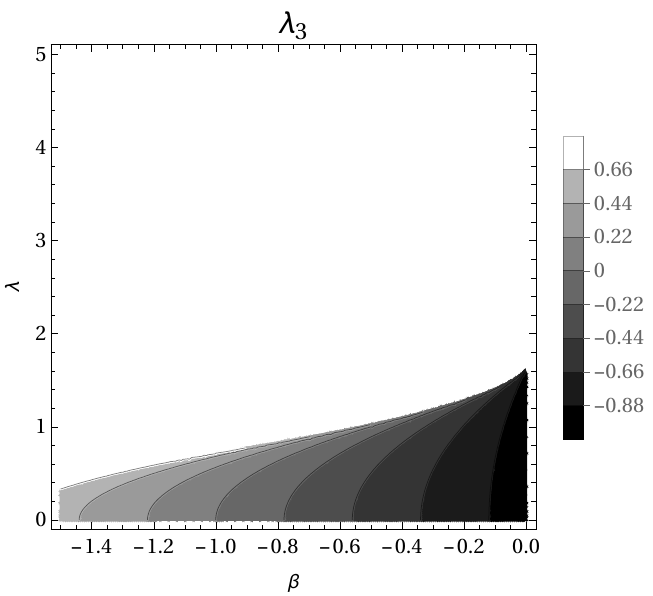}
    \end{minipage}
    \hfill
     \begin{minipage}[t]{0.325\textwidth}
        \centering
        \includegraphics[width=\textwidth]{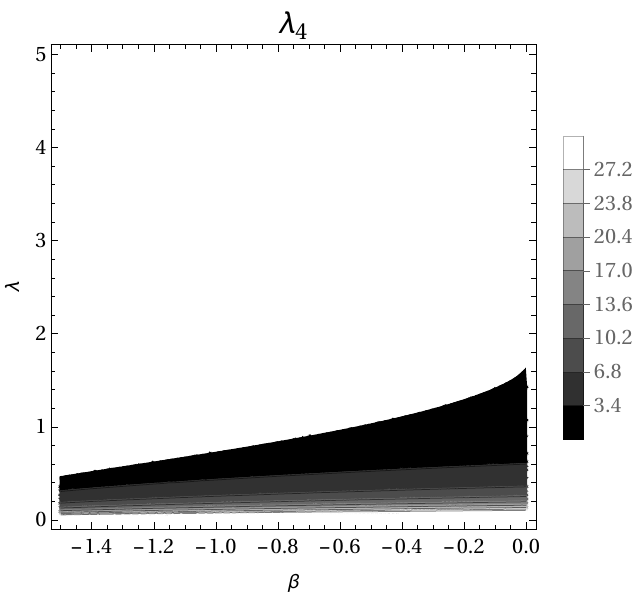}
    \end{minipage}
    
    \vspace{0.2cm}
    
    \begin{minipage}[t]{0.325\textwidth}
        \centering
        \includegraphics[width=\textwidth]{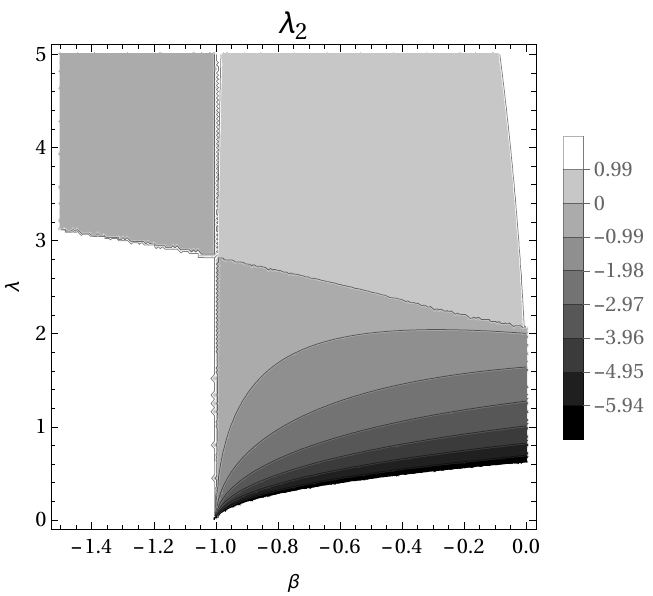} 
    \end{minipage}
    \hfill
     \begin{minipage}[t]{0.325\textwidth}
        \centering
        \includegraphics[width=\textwidth]{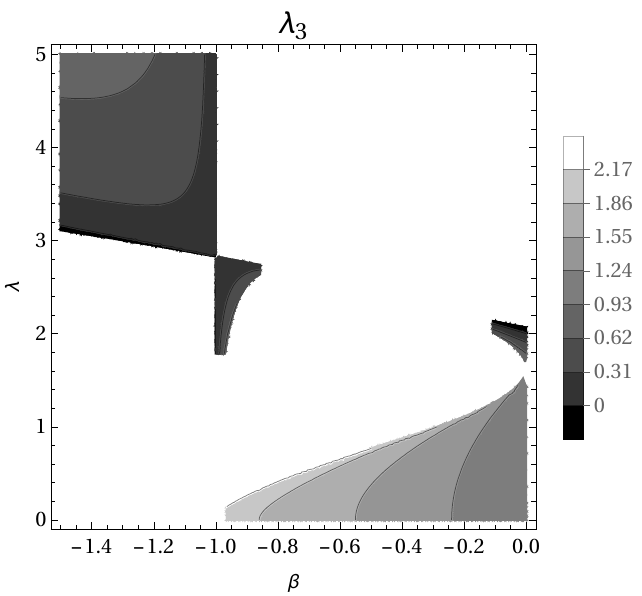}
    \end{minipage}%
    \hfill
    \begin{minipage}[t]{0.325\textwidth}
        \centering
        \includegraphics[width=\textwidth]{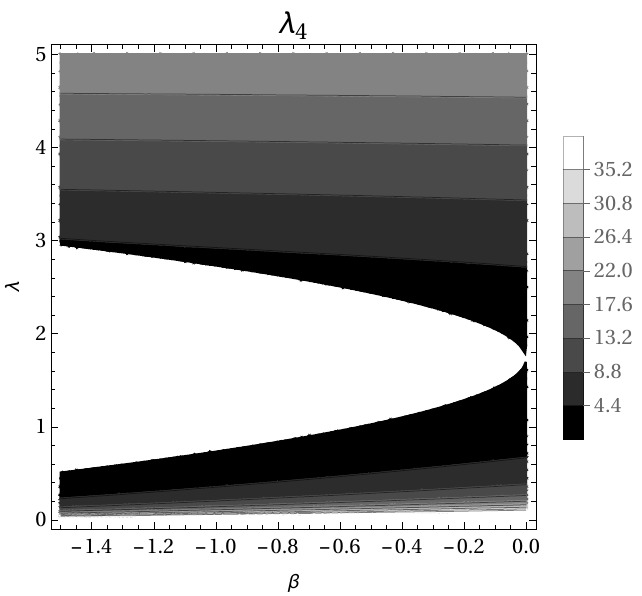}
    \end{minipage}
    \caption{Here, the first row shows the plots for the eigenvalues of $P_{mixed_{1}}$ and second row shows the plots for $P_{mixed_{2}}$ to show the stability for the two fixed points.}
    \label{figure1}
\end{figure}

\noindent Let us first consider the stability of the fixed point $P_{{mixed}_1}$. In the first row of Fig. (\ref{figure1}), we plot the eigenvalues $(\lambda_2,\lambda_3,\lambda_4)$ for the fixed point $P_{{mixed}_1}$ as a contour plot of $\lambda$ and $\beta$ within the range given in Eq. (\ref{rangem2}). The figure shows that the eigenvalue $\lambda_2$ is always negative. On the other hand, $\lambda_3$ exhibits both positive and negative values depending on the values of $\beta$ and $\lambda$, while $\lambda_4$ remains positive within the given range. The behavior of the eigenvalues indicates that $P_{{mixed}_1}$ is a saddle point.

Similarly for $P_{{mixed}_2}$, we plot the eigenvalues $(\lambda_2,\lambda_3,\lambda_4)$ in the second row of Fig. (\ref{figure1}). In this case, the eigenvalue $\lambda_2$ can take both negative and positive values, the eigenvalue $\lambda_3$ is entirely positive while $\lambda_4$ is completely negative. Based on the behaviour of the eigenvalues, it can be concluded that this point is also a saddle point.

\begin{figure}[H]
    \begin{minipage}[t]{0.37\textwidth}
        \centering
        \includegraphics[width=\textwidth]{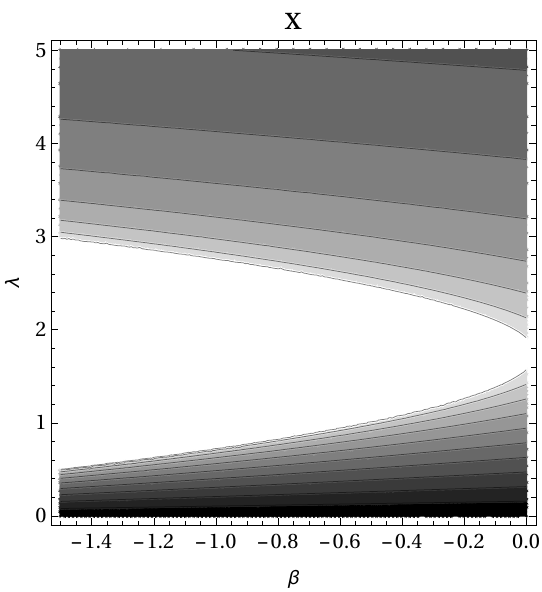}
    \end{minipage}%
    \hspace{1.7cm}
    \begin{minipage}[t]{0.37\textwidth}
        \centering
        \includegraphics[width=\textwidth]{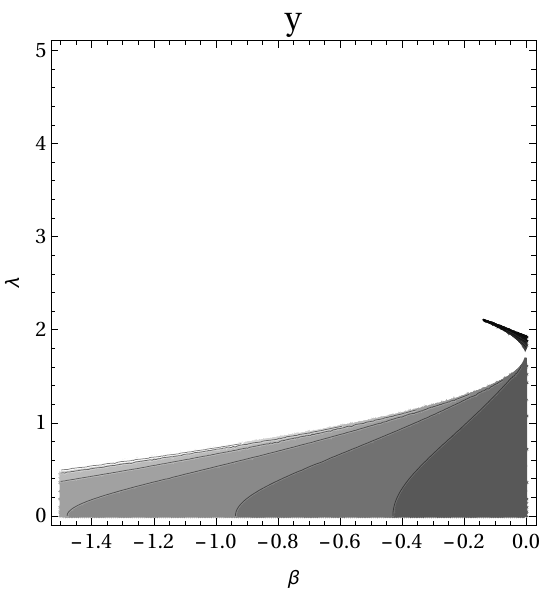}
    \end{minipage}

    \vspace{0.1cm}

    \begin{minipage}[t]{0.435\textwidth}
        \centering
        \includegraphics[width=\textwidth]{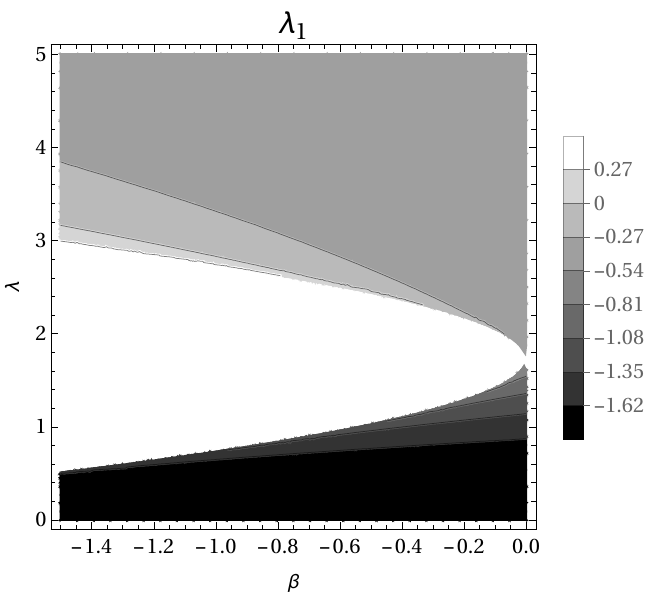} 
    \end{minipage}
    \hspace{0.5cm}
     \begin{minipage}[t]{0.435\textwidth}
        \centering
        \includegraphics[width=\textwidth]{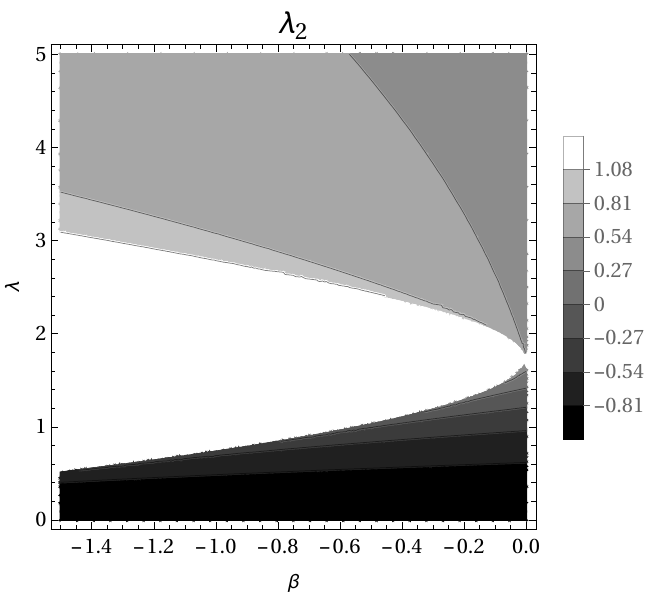}
    \end{minipage}

    \vspace{0.1cm}

    \begin{minipage}[t]{0.435\textwidth}
        \centering
        \includegraphics[width=\textwidth]{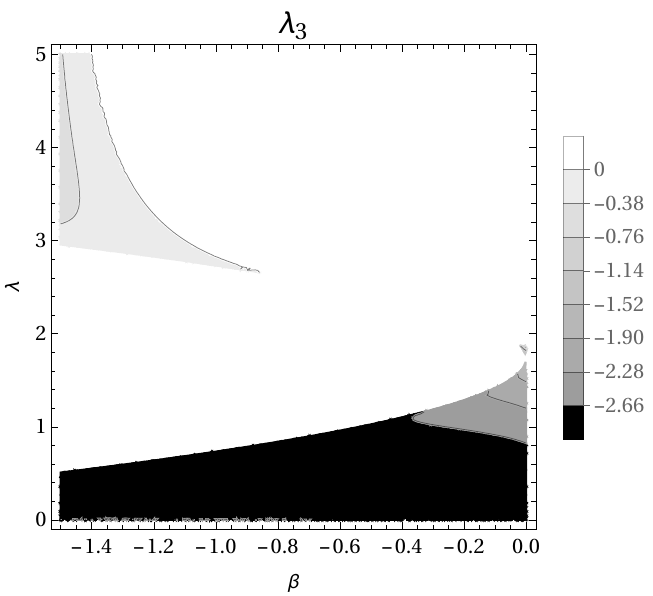}
    \end{minipage}
    \hspace{0.5cm}
    \begin{minipage}[t]{0.435\textwidth}
        \centering
        \includegraphics[width=\textwidth]{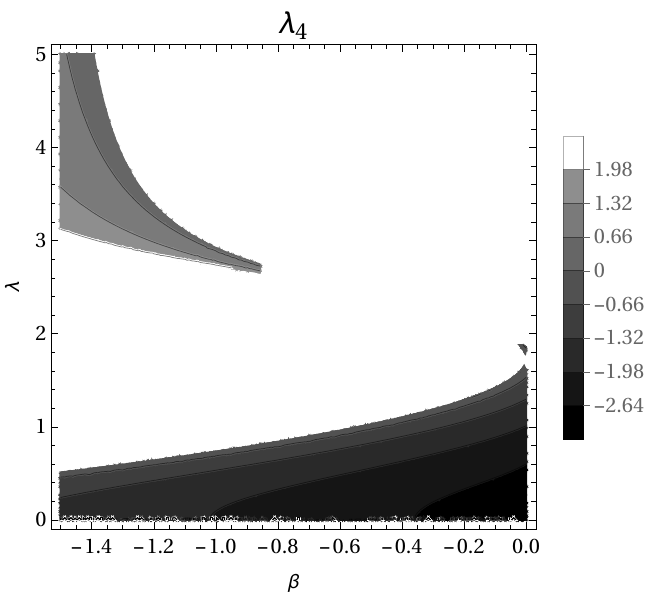}
    \end{minipage}
    \caption{Plot for the fixed point $P_{m\phi_{2}}$ and its eigenvalues.}
    \label{figure2}
\end{figure}

\noindent For the fixed points $P_{{m\phi}_2}$ and $P_{{m\phi}_3}$, the question of their existence itself is a complicated issue. Let us discuss $P_{{m\phi}_2}$ first. For this point to exist both $x$ and $y$ should be real  which will happen if and only if  $P_1$, $P_2$ given in Eq. (\ref{p1}), (\ref{p2}) have real roots. The plot $P_1$, $P_2$ as function of $\beta$ and $\lambda$ are given in first row of Fig. (\ref{figure2}). The parameter space where this point exists is given by the overlap of the shaded regions of those two plots. The second and the third rows of Fig. (\ref{figure2}) shows the contour plot of the eigenvalues  ($\lambda_1,\lambda_2,\lambda_3,\lambda_4$) associated with this fixed point. We need to consider only those regions of the eigenvalue plots which overlap with the parameter space where both $x$ and $y$ exist.  As can be seen easily from Fig. (\ref{figure2}), in that region, all the eigenvalues are negative indicating that $P_{{m\phi}_2}$ is a stable fixed point. 

\begin{figure}[H]
    \begin{minipage}[t]{0.37\textwidth}
        \centering
        \includegraphics[width=\textwidth]{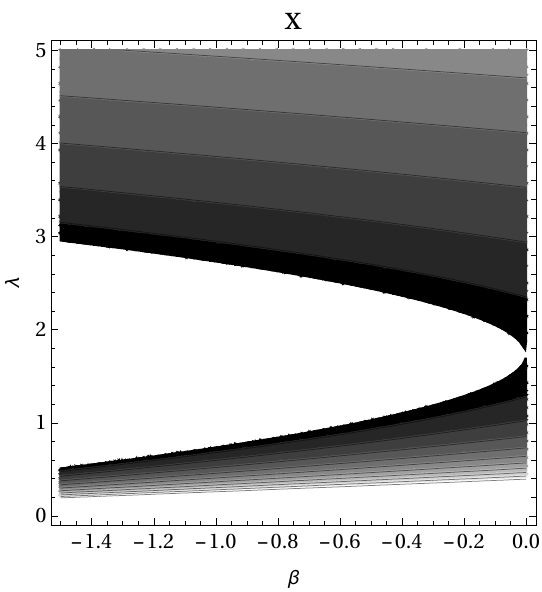}
    \end{minipage}%
    \hspace{1.7cm}
    \begin{minipage}[t]{0.37\textwidth}
        \centering
        \includegraphics[width=\textwidth]{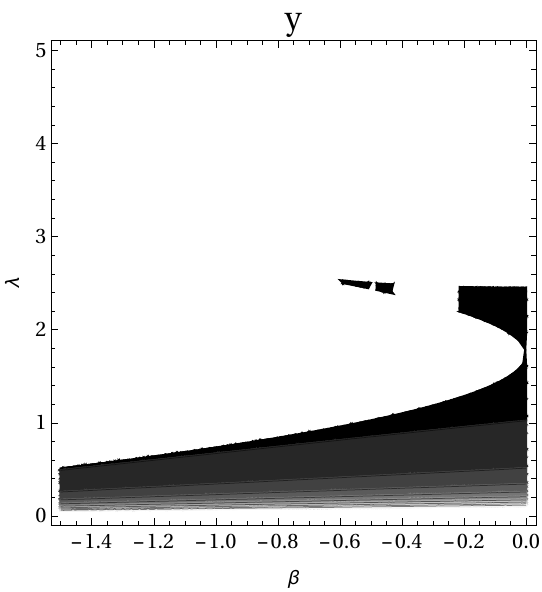}
    \end{minipage}

    \vspace{0.1cm}

    \begin{minipage}[t]{0.435\textwidth}
        \centering
        \includegraphics[width=\textwidth]{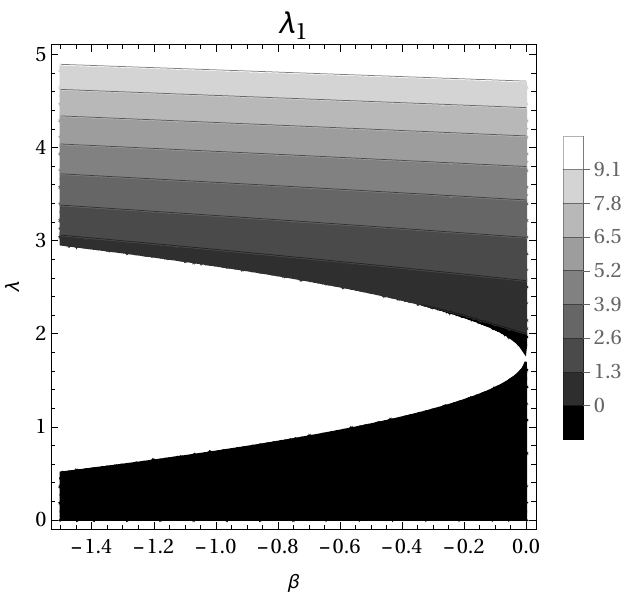} 
    \end{minipage}
    \hspace{0.5cm}
     \begin{minipage}[t]{0.435\textwidth}
        \centering
        \includegraphics[width=\textwidth]{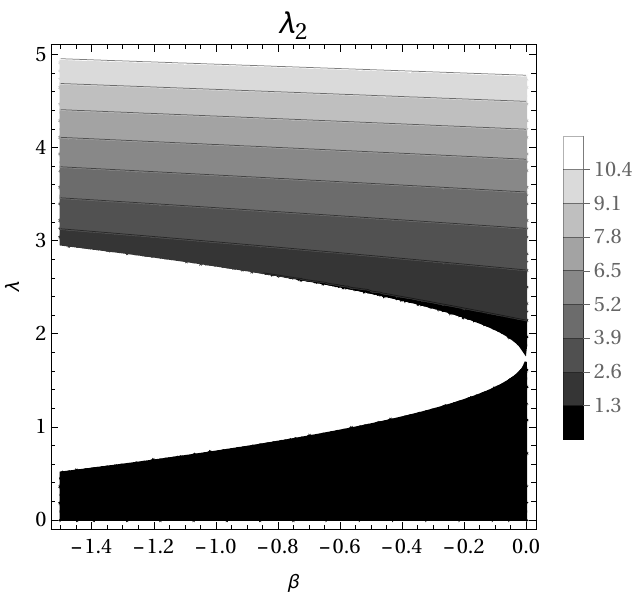}
    \end{minipage}

    \vspace{0.1cm}

    \begin{minipage}[t]{0.435\textwidth}
        \centering
        \includegraphics[width=\textwidth]{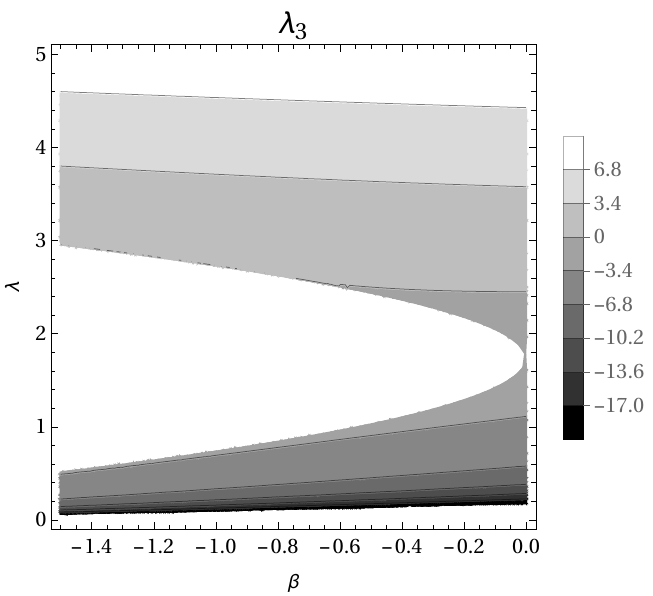} 
    \end{minipage}
    \hspace{0.5cm}
    \begin{minipage}[t]{0.435\textwidth}
        \centering
        \includegraphics[width=\textwidth]{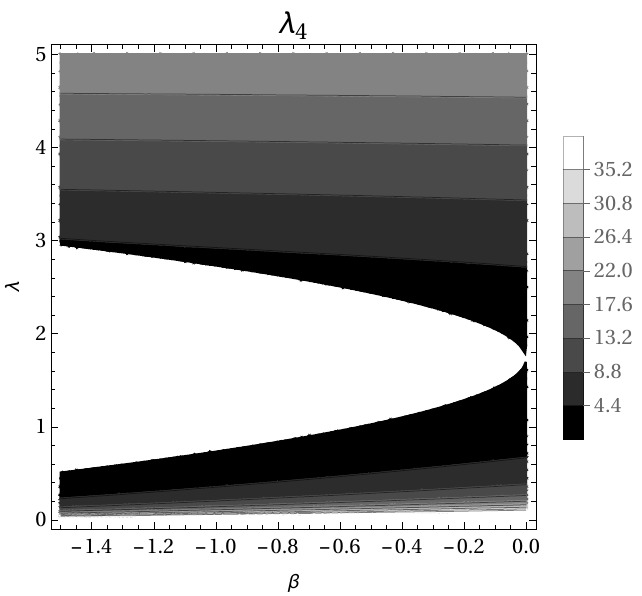} 
    \end{minipage}
    \caption{Plot for the fixed point $P_{m\phi_{3}}$ and its eigenvalues.}
    \label{figure3}
\end{figure}

\noindent From a similar analysis given in Fig. (\ref{figure3}), we can conclude that
$P_{{m\phi}_3}$ is a saddle point. In Fig. (\ref{figure3}), the first row plots the parameter space in which $P_3$ and $P_4$ have real roots, while the next two rows are contour plots of the eigenvalues. \\

\noindent With the above analysis in mind let us describe the fixed points. The cosmological parameters corresponding to each of the fixed points are given in Table (\ref{table:6}),
\begin{enumerate}
    \item Kination Phase ($P_{{m\phi}_1}^{\pm}$): As discussed above, we will choose parameters, $\beta$ and $\lambda$ such that it becomes an unstable fixed point. $P_{{m\phi}_1}^-$ is unstable for all ranges while the range for which $P_{{m\phi}_1}^+$ becomes unstable is given in Eq. (\ref{rangem2}). Note that although it is an unstable fixed point, both DM and DE densities contribute. Also note that  if we take $\beta=-\frac{3}{2}$, this point mimics the matter domination epoch of the universe and merges with the point $P_m$.
      
    \item Matter domination ($P_m$): This is a saddle point where $\Omega_m = 1$ and $\omega_{eff} =0$. This indicates a matter dominated phase.

    \item Radiation domination ($P_r$): This is a saddle point where
    $\Omega_r =1$ and $\omega_{eff}=\frac{1}{3}$ indicating a radiation dominated regime.

    \item Curvature domination ($P_c$): This again is a saddle point where $\Omega_k = 1$ which indicates a curvature dominated phase.
    
\begin{table}[H]
\addtolength{\tabcolsep}{3.5pt}
\renewcommand{\arraystretch}{2.2}
\centering
\begin{tabular}{| c | c | c | c | c | c | c | c | c | c | c | c |} 
 \hline
Point & $\Omega_m$ & $\Omega_r$ & $\Omega_\phi$ & $\Omega_k$ & 
$\omega_{eff}$ & $q$ & $\omega_{m_2}$ \\ 
\hline\hline
$P_{{m \phi}_1}^\pm$ & $-\frac{2 \beta}{3}$ & $0$ & $1+\frac{2 \beta}{3}$ & $0$ & $1+\frac{2 \beta}{3}$ & $2+\beta$ & $1+\frac{2 \beta}{3}$ \\ [0.5ex]
\hline

$P_m$ & $1$ & $0$ & $0$ & $0$ & $0$ & $\frac{1}{2} $ & $0$\\
\hline

$P_r$ & $0$ & $1$ & $0$ & $0$ & $\frac{1}{3}$ & $1$ & Indeterminate\\ [0.5ex]
\hline

$P_c$ & $0$ & $0$ & $0$ & $1$ & $-\frac{1}{3}$ & $0$ & Indeterminate\\ [0.5ex]
\hline

$P_{mixed_1}$ & $\frac{4\beta}{3\lambda^2} $ & $0$ & $\frac{2(3+\beta)}{3\lambda^2}$ & $1-\frac{2(1+\beta)}{\lambda^2}$ & $-\frac{1}{3}$ & $0$ & $-\frac{1}{3}$ \\  [0.5ex]
\hline

$P_{mixed_2}$ & $-\frac{16\beta}{3\lambda^2}$ & $1+\frac{4(\beta-1)}{\lambda^2}$ & $\frac{4(3+\beta)}{3\lambda^2}$ & $0$ & $\frac{1}{3}$ & $1$ & $\frac{1}{3}$ \\  [0.5ex]
\hline

$P_{{m \phi}_2}$ & $\Omega_{m_1}$ & $0$ & $\Omega_{\phi_1}$ & $0$ & $\omega_{{eff}_1}$ & $q_1$ & $\Tilde{\omega}_{1}$\\ [0.5ex]
\hline

$P_{{m \phi}_3}$ & $\Omega_{m_2}$ & $0$ & $\Omega_{\phi_2}$ & $0$ & $\omega_{{eff}_2}$ & $q_2$ & $\Tilde{\omega}_{2}$ \\ [0.5ex]

\hline
\end{tabular}
\caption{List of the cosmological parameters for Model $2$.} 
\label{table:6}
\end{table}

    \item Mixed phase ($P_{mixed_1}$ and $P_{mixed_2}$): These are two saddle points which exist due to our specific form of interaction and choice of potential. In $P_{{mixed}_1}$, matter, scalar field and curvature contribute to the energy density but it mimics the behaviour of curvature dominated phase. Similarly in $P_{{mixed}_2}$, matter, scalar field and radiation contribute to the energy density but mimics the radiation dominated phase.
    
     \item Matter Scaling ($P_{{m\phi}_3}$): This is a saddle point which exist because of the specific form of interaction and the potential chosen. The evolution of the universe is driven by both the scalar field and matter. The $w_{eff}$ for this point will be discussed later. 

     \item Dark Matter Scaling ($P_{{m\phi}_2}$): Again, based on the analysis above, we have this as a stable fixed point. The exact behaviour of $w_{eff}$ is discussed below.    
\end{enumerate}
From Table (\ref{table:5}), we see that the universe starts from a kination epoch $P_{{m\phi}_1}$, goes through phases of matter domination $P_m$, radiation domination $P_r$, curvature domination $P_c$ and other intermediate saddle phases $P_{{mixed}_1}$, $P_{{mixed}_2}$, $P_{{m\phi}_3}$ and finally reaches $P_{{m\phi}_2}$ which is a stable point. 

\begin{figure}[H]
    \begin{minipage}[t]{0.45\textwidth}
        \centering
        \includegraphics[width=\textwidth]{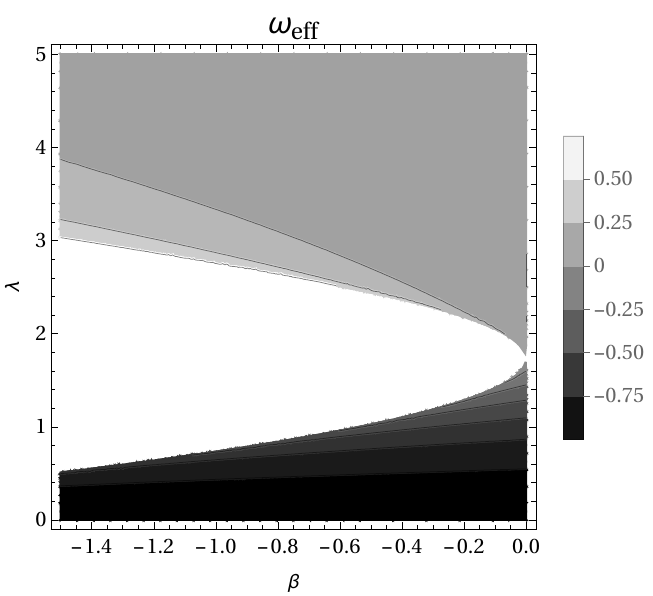} 
    \end{minipage}
    \hspace{0.5cm}
    \begin{minipage}[t]{0.45\textwidth}
        \centering
        \includegraphics[width=\textwidth]{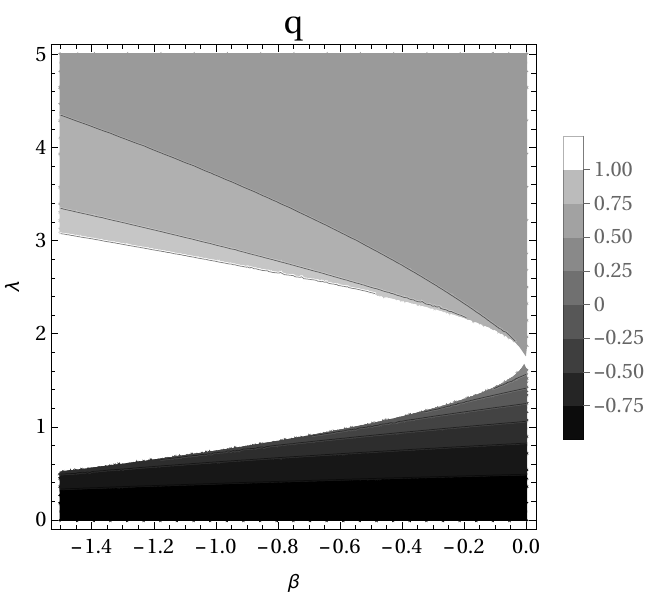}
    \end{minipage}
    
    \vspace{0.2cm}
    
    \begin{minipage}[t]{0.45\textwidth}
        \centering
        \includegraphics[width=\textwidth]{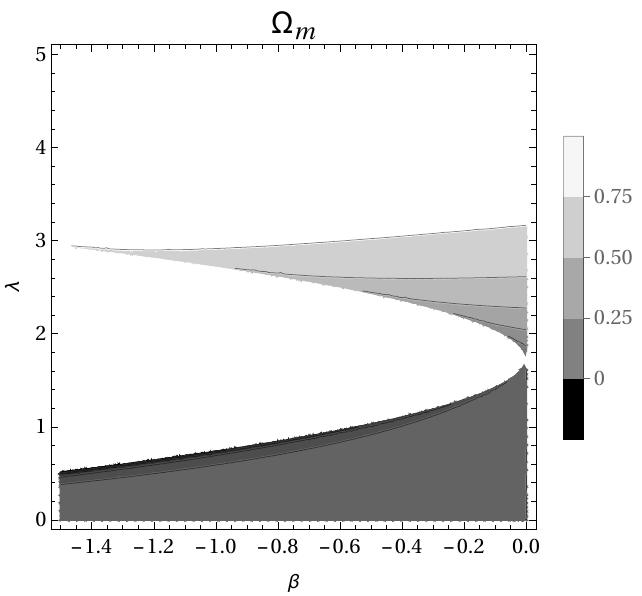} 
    \end{minipage}
    \hspace{0.5cm}
    \begin{minipage}[t]{0.45\textwidth}
        \centering
        \includegraphics[width=\textwidth]{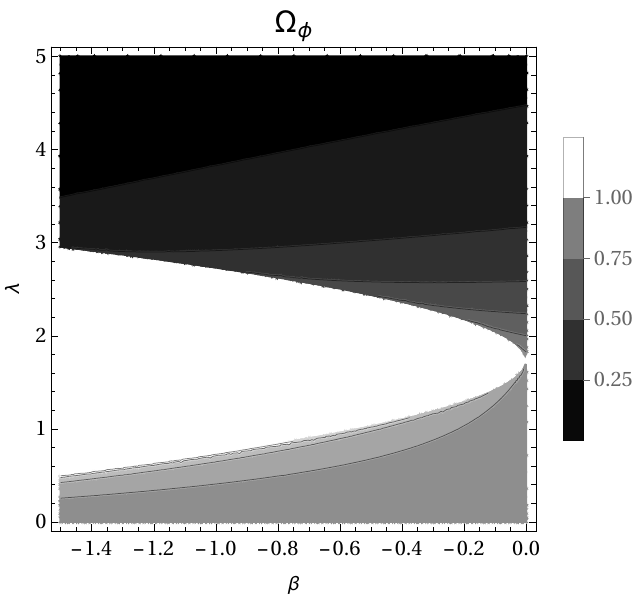} 
    \end{minipage}
    \caption{Plot for the different cosmological parameters $\omega_{eff}$, q, $\Omega_m$ and $\Omega_{\phi}$ for the exponential potential for the fixed point $P_{{m\phi}_2}$.}
    \label{figure4}
\end{figure}

Let us analyze the features $P_{{m \phi}_2}$ in a bit more detail. The effective EOS, deceleration parameter, the relative energy densities and EOS for DM for this fixed points are: 
\begin{eqnarray}
    &&\omega_{{eff}_1} = \frac{-3+\lambda^2+\beta -\delta}{6}, ~ ~ ~ q_1 = \frac{-1+\lambda^2+\beta -\delta}{4}, 
    \nonumber \\
    &&\Omega_{m_1} = \frac{3-\beta }{6}+\frac{(3+\beta ) \left(-3-\beta +\delta\right)}{6 \lambda^2}, ~ ~ ~ \Omega_{\phi_1} = \frac{(3+\beta ) \left(3+\lambda^2+\beta -\delta\right)}{6 \lambda^2} \nonumber \\
    &&\Tilde{\omega}_{1} = -\frac{\beta \left( 3+\lambda^2 +\beta-\delta \right)^2}{6\Bigl( -\lambda^2(-3+\beta) + (3+\beta) (-3-\beta+\delta) \Bigr)} 
\end{eqnarray}
Again, since a algebraic analysis is difficult, we will determine the properties from the contour plots of these quantities as functions of $\lambda$ and $\beta$ given in Fig. (\ref{figure4}). The physically relevant regions of Fig. (\ref{figure4}) are where both $x$ and $y$ coordinates of $P_{{m \phi}_2}$ exist (given in Fig. (\ref{figure2})). 
 
We observe that both the $\omega_{eff}$ and $q$ range from $-0.25$ to $-1$ for the same range of $\beta$ and $\lambda$ from the overlap region of $x$ and $y$. This negative $\omega_{eff}$ and $q$ point towards the late time acceleration of the universe. Similarly, the relative energy density $\Omega_m$ is from $0$ to $0.25$ and $\Omega_\phi$ is from $0.75$ to $1$ for the same range of $\beta$ and $\lambda$ indicating that both DM and DE drive this acceleration. The behaviour of $\omega_{m_2}$ is discussed in Appendix (\ref{appendix2}).

\begin{figure}[H]
    \begin{minipage}[t]{0.45\textwidth}
        \centering
        \includegraphics[width=\textwidth]{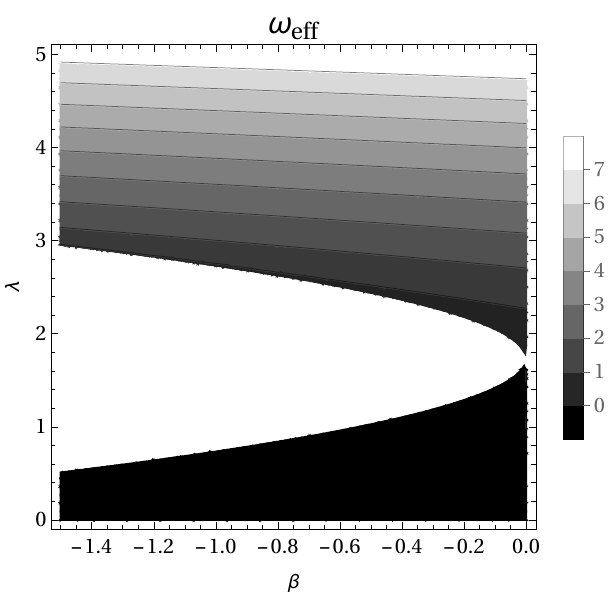} 
    \end{minipage}
    \hspace{0.5cm}
    \begin{minipage}[t]{0.45\textwidth}
        \centering
        \includegraphics[width=\textwidth]{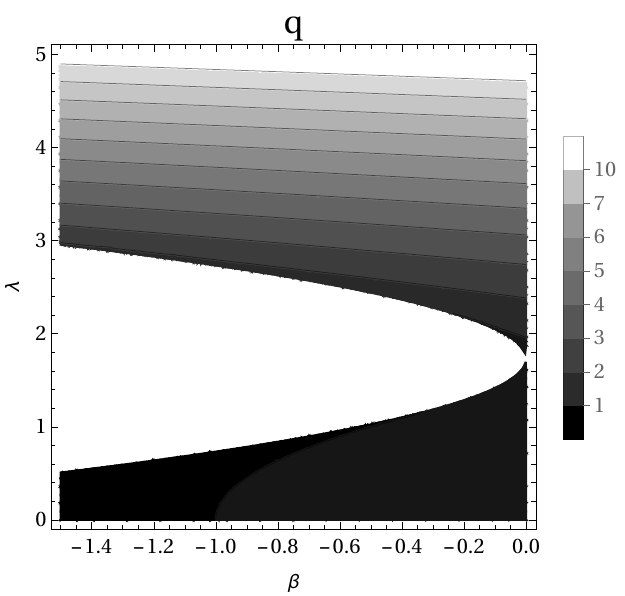}
    \end{minipage}
    
    \vspace{0.2cm}
    
     \begin{minipage}[t]{0.45\textwidth}
        \centering
        \includegraphics[width=\textwidth]{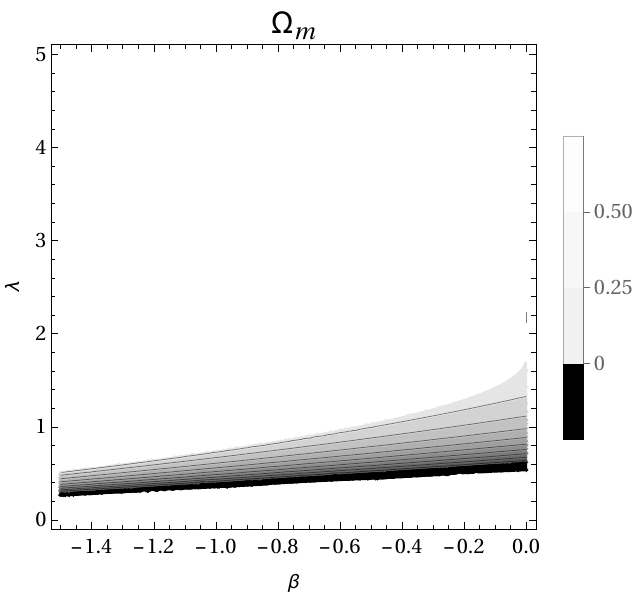}         
    \end{minipage}
    \hspace{0.5cm}
    \begin{minipage}[t]{0.45\textwidth}
        \centering
        \includegraphics[width=\textwidth]{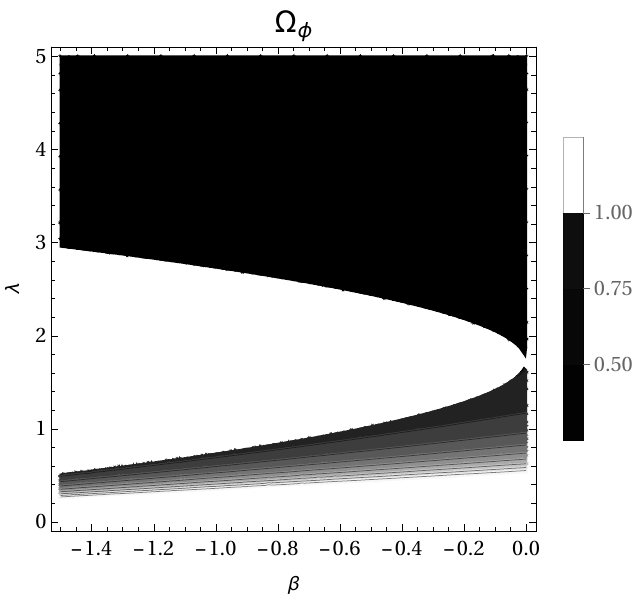} 
    \end{minipage}
    \caption{Plot for the different cosmological parameters $\omega_{eff}$, q, $\Omega_m$ and $\Omega_{\phi}$ for the exponential potential for the fixed point $P_{{m\phi}_3}$.}
    \label{figure5}
\end{figure}

Similiarly, the effective EOS, deceleration parameter, relative energy densities and the EOS for DM for $P_{{m \phi}_3}$ are: 
\begin{eqnarray}
    &&\omega_{{eff}_2} = \frac{-3+\lambda^2+\beta +\delta}{6}, ~ ~ ~ q_2 = \frac{-1+\lambda^2+\beta +\delta}{4}
    \nonumber \\
    &&\Omega_{m_2} = \frac{3-\beta }{6}-\frac{(3+\beta ) \left(3+\beta +\delta\right)}{6 \lambda^2}, ~ ~ ~ \Omega_{\phi_2} = \frac{(3+\beta ) \left(3+\lambda^2+\beta +\delta\right)}{6 \lambda^2} \nonumber \\
    &&\Tilde{\omega}_{2} = \frac{\beta \left( 3+\lambda^2 +\beta+\delta \right)^2}{6\Bigl(\lambda^2(-3+\beta) + (3+\beta) (3+\beta+\delta) \Bigr)} 
\end{eqnarray}

From Fig. (\ref{figure5}), we see that the $\omega_{eff}$ and $q$ are positive in the overlap region while the $\Omega_m$ varies from $0$ to $0.5$ and $\Omega_\phi$ varies from $0.5$ to $1$ for the fixed point $P_{{m\phi}_3}$. Again the discussion about $\omega_{m_2}$ is given in Appendix (\ref{appendix2}).

\begin{figure}[H]
    \begin{minipage}[t]{0.45\textwidth}
        \centering
        \includegraphics[width=\textwidth]{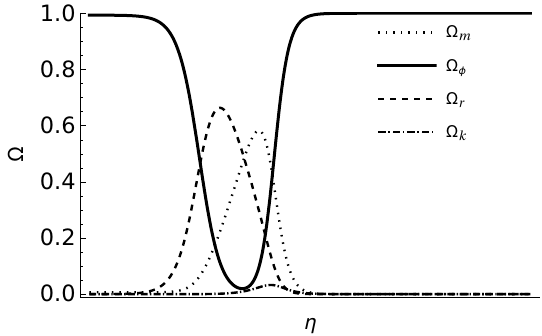} 
    \end{minipage}
    \hspace{0.5cm}
    \begin{minipage}[t]{0.45\textwidth}
        \centering
        \includegraphics[width=\textwidth]{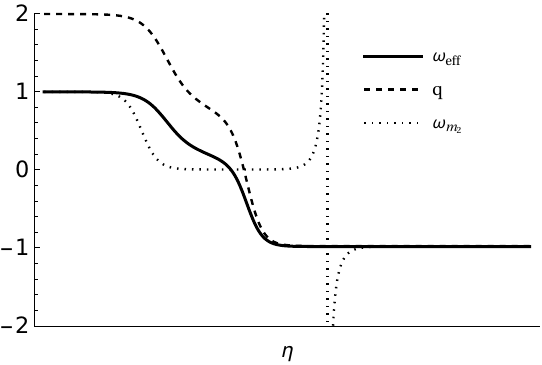}
    \end{minipage}
    
    \vspace{0.6cm}
    
    \begin{minipage}[t]{0.45\textwidth}
        \centering
        \includegraphics[width=\textwidth]{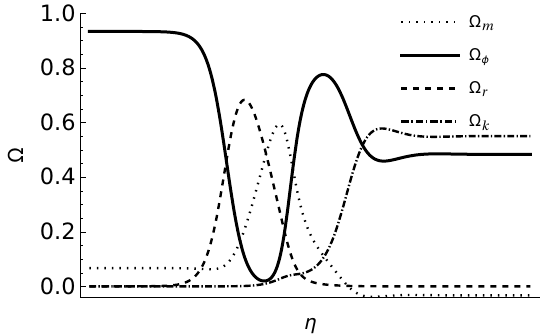} 
    \end{minipage}
    \hspace{0.5cm}
    \begin{minipage}[t]{0.45\textwidth}
        \centering
        \includegraphics[width=\textwidth]{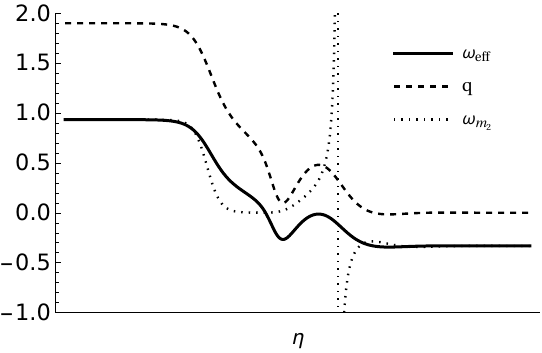} 
    \end{minipage}
    \caption{Evolution plot for the different cosmological parameters. Here, the first row represents the plots for $\lambda=0.2$ and $\beta=-0.01$ while the second row represents the plots for $\lambda=2$ and $\beta=-0.1$.}
    \label{figuree2}
\end{figure}
\noindent

The evolution plot for different cosmological parameters is given in Fig. (\ref{figuree2}).
The first row of Fig. (\ref{figuree2}) is plotted for small values of $\beta (=-0.01)$ and $\lambda (=0.2)$. This plot shows that the universe starts evolving due to field domination showed by thick black line, then radiation domination occurs and after that there is matter domination and again at late times, field starts dominating. Note that the contribution arising from curvature is negligible for small values of $\lambda$ and $\beta$. Then, we have plotted EOS and $q$ for the same $\lambda$ and $\beta$, we see that EOS and $q$ are negative for late times indicating this is a viable model for our universe.

In the second row of Fig. (\ref{figuree2}), we again have plotted the evolution for larger values of $\lambda$ and $\beta$. We have taken $\lambda =2$ and $\beta=-0.1$. We see that now the universe starts from field domination then there is radiation domination, matter domination and again field domination but after that there is contribution from curvature as well. Since $q$ is close to $0$ at late times even though $\omega_{eff}<0$, this may not be a good description of our universe. 

In both these plots $\omega_{m_2}$ is plotted by dotted black line. It is behaving like stiff matter ($\omega_{m_2} =1$) at early times, then becomes $0$ thus behaving as dust and becomes negative at late times which is like exotic DM. As discussed above, at certain points $\omega_{m_2}$ blows up and thus cannot be used to describe the nature of the DM at those points. The effective exotic nature of DM is more apparent for larger values of $\lambda$ and $\beta$ (first figure of bottom row of Fig. (\ref{figuree2})). 

\noindent Let us look at a few interesting sub-case of our model:

\subsubsection{No Radiation}
An invariant subspace of our model is obtained when the radiation effects are switched off by setting $u=0$. The behaviour of fixed points, their stability and the cosmological parameters corresponding to these points are given in Table (\ref{table:8}). This time the points $P_r$ and $P_{{mixed}_2}$ where $\omega_{eff} = \frac{1}{3}$ disappear as expected. 

\begin{table}[H]
\addtolength{\tabcolsep}{-1.5pt}
\renewcommand{\arraystretch}{2.2}
\centering
\begin{tabular}{| c | c | c | c | c | c | c | c | c | c | c | c |} 
 \hline
Point & Stability & $\omega_{eff}$ & $\Omega_m$ &  $\Omega_\phi$ & $\Omega_k$ & q & $\omega_{m_2}$\\ 
\hline\hline
$P_{{m \phi}_1}^\pm$ & $P^+_{{m\phi}_1}$: \multirow{2}{*}{\(\left\{\begin{array}{c} \\ \\ \end{array}\right.\)} Unstable for $\lambda<\frac{6+2\beta}{\sqrt{6+4\beta}}$ & $1+\frac{2 \beta}{3}$ & $-\frac{2 \beta}{3}$ &  $1+\frac{2 \beta}{3}$ & $0$ & $2+\beta$ & $1+\frac{2 \beta}{3}$  \\ 

&  ~ ~ ~ ~ ~ ~ ~ Saddle for $\lambda \ge \frac{6+2\beta}{\sqrt{6+4\beta}}$ & & & & & & \\
 
& $P^-_{{m\phi}_1}$: Fully unstable  & & & & & & \\
[0.5ex]
\hline

$P_m$ & Saddle & $0$ & $1$ & $0$ & $0$ & $\frac{1}{2}$ & $0$ \\
\hline

$P_c$ & Saddle & $-\frac{1}{3}$ & $0$ & $0$ & $1$ & $0$ & Indeterminate \\ [0.5ex]
\hline

$P_{mixed_1}$ & Saddle (See Fig. (\ref{figure1}) first row) & $-\frac{1}{3}$ & $\frac{4\beta}{3\lambda^2}$ & $\frac{2(3+\beta)}{3\lambda^2}$ & $1-\frac{2(1+\beta)}{\lambda^2}$  & $0$ & $-\frac{1}{3}$ \\  [0.5ex]
\hline

$P_{{m \phi}_2}$ & Stable (See Fig. (\ref{figure2}) ($\lambda_2, \lambda_3, \lambda_4$)) & $\omega_{{eff}_1}$ & $\Omega_{m_1}$ & $\Omega_{\phi_1}$ & $0$  & $q_1$ & $\Tilde{\omega}_{1}$\\ [0.5ex]
\hline

$P_{{m \phi}_3}$ & Saddle (See Fig. (\ref{figure3}) ($\lambda_2, \lambda_3, \lambda_4$)) & $\omega_{{eff}_2}$ & $\Omega_{m_2}$  & $\Omega_{\phi_2}$ & $0$ & $q_2$ & $\Tilde{\omega}_{2}$\\ [0.5ex]

\hline
\end{tabular}
\caption{Dynamics of the fixed points and their stability when we switch off the radiation.}
\label{table:8}
\end{table}

\subsubsection{Flat universe:~$k = 0$}
The invariant subspace of our model where the curvature effects are switched off is obtained by setting $z=0$. The fixed points, their stabilty and the cosmological parameters corresponding to these points are given in Table (\ref{table:7}). Unsurprisingly, the behaviour of the fixed points change and the points $P_c$ and $P_{{mixed}_1}$ where $\omega_{eff} = -\frac{1}{3}$ disappear.

\begin{table}[H]
\addtolength{\tabcolsep}{-1.5pt}
\renewcommand{\arraystretch}{2.2}
\centering
\begin{tabular}{| c | c | c | c | c | c | c | c | c | c | c | c |} 
 \hline
Point & Stability & $\omega_{eff}$ & $\Omega_m$ & $\Omega_r$ & $\Omega_\phi$ & q & $\omega_{m_2}$\\ 
\hline\hline
$P_{{m \phi}_1}^\pm$ & $P^+_{{m\phi}_1}$: \multirow{2}{*}{\(\left\{\begin{array}{c} \\ \\ \end{array}\right.\)} Unstable for $\lambda<\frac{6+2\beta}{\sqrt{6+4\beta}}$ & $1+\frac{2 \beta}{3}$ & $-\frac{2 \beta}{3}$ & $0$ & $1+\frac{2 \beta}{3}$ & $2+\beta$ & $1+\frac{2 \beta}{3}$ \\ 

&  ~ ~ ~ ~ ~ ~ ~ Saddle for $\lambda \ge \frac{6+2\beta}{\sqrt{6+4\beta}}$ & & & & & & \\
 
& $P^-_{{m\phi}_1}$: Fully unstable  & & & & & & \\
[0.5ex]
\hline

$P_m$ & Saddle & $0$ & $1$ & $0$ & $0$ & $\frac{1}{2}$ & $0$\\
\hline

$P_r$ & Saddle for $\beta \ge -1$ & $\frac{1}{3}$ & $0$ & $1$ & $0$ & $1$ & Indeterminate\\ [0.5ex]
\hline

$P_{mixed_2}$ & Saddle (See Fig. (\ref{figure1}) second row) & $\frac{1}{3}$ & $-\frac{16\beta}{3\lambda^2}$ & $1+\frac{4(\beta-1)}{\lambda^2}$ & $\frac{4(3+\beta)}{3\lambda^2}$ & $1$ & $\frac{1}{3}$ \\  [0.5ex]
\hline

$P_{{m \phi}_2}$ & Stable (See Fig. (\ref{figure2}) ($\lambda_1, \lambda_3, \lambda_4$)) & $\omega_{{eff}_1}$ & $\Omega_{m_1}$ & $0$ & $\Omega_{\phi_1}$ & $q_1$ & $\Tilde{\omega}_{1}$\\ [0.5ex]
\hline

$P_{{m \phi}_3}$ & Saddle (See Fig. (\ref{figure3}) ($\lambda_1, \lambda_3, \lambda_4$)) & $\omega_{{eff}_2}$ & $\Omega_{m_2}$ & $0$ & $\Omega_{\phi_2}$ & $q_2$ & $\Tilde{\omega}_{2}$\\ [0.5ex]

\hline
\end{tabular}
\caption{List of fixed points and their stability for Model $2$ for $k=0$.}
\label{table:7}
\end{table}

\begin{figure}[H]
    \begin{minipage}[t]{0.45\textwidth}
        \centering
        \includegraphics[width=\textwidth]{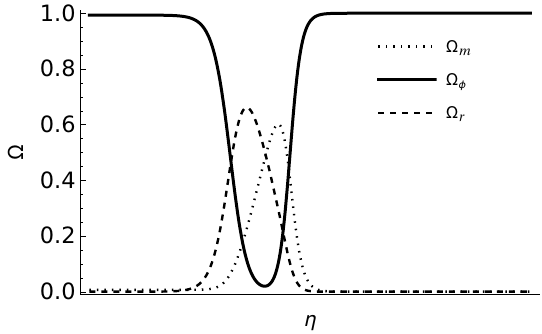} 
    \end{minipage}
    \hspace{0.5cm}
    \begin{minipage}[t]{0.45\textwidth}
        \centering
        \includegraphics[width=\textwidth]{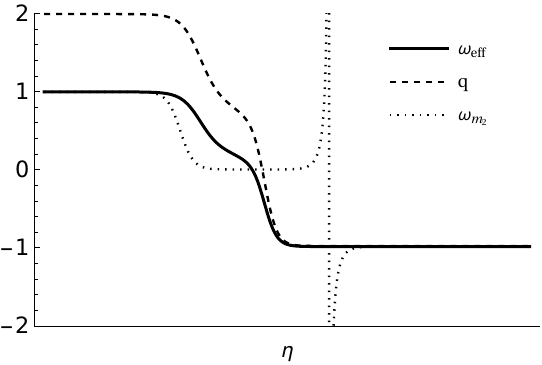}
    \end{minipage}
    \caption{Evolution plot for the different cosmological parameters for $\lambda=0.2$ and $\beta=-0.01$.}
    \label{figureevolz2}
\end{figure}    
The evolution plot for the flat case is given in Figure (\ref{figureevolz2}). 
We see there is late time acceleration driven by a interacting DE and exotic DM. Again the effective EOS of DM $\omega_{m_2}$ varies over the course of evolution. Here it starts off as stiff matter, then transitions to dust and then shows exotic DM behaviour at late times. (See Fig. (\ref{figuree2}) for $k=-1$, and Fig. (\ref{figureevolz2}) for $k=0$). This is similar to the observations made in \cite{deogharia2021generalized,roy2024dynamicalstabilityphasespace,Hussain_2023}. Unlike Model $1$ the transition as described by $\omega_{m_2}$ is not smooth because at certain points in evolution history $\omega_{m_2}$ becomes unbounded, although the phase space constraint is always satisfied.

\section{Conclusion:} \label{section4} 
 
Let us briefly recap our results. We model all the matter content (including DM) as fluids and the DE as a minimally coupled canonical scalar field. 
 Different models will correspond to different choices of the potential and the form of the interaction between DM and DE. We explored two such interacting DM-DE models. 
\begin{itemize}
    \item {\bf{Model 1}}: $  V = V_0 e^{-\lambda \phi} $ and  $  Q_I =  \sqrt{\frac{2}{3}} \kappa \beta \Dot{\phi} \rho_m$
    \item {\bf{Model 2}}: $  V = V_0 e^{-\lambda \phi} $ and $ Q_I =  \beta H \dot{\phi}^2$
\end{itemize}
Motivated by the observation of a DE dominated universe at late times, we take $\beta$ to be negative so that energy flows from DM to DE.

A dynamical systems analysis is carried out and the fixed points and their stability analyzed. Demanding the existence of an unstable fixed point (or an unstable spiral) which will represent the early universe and a stable fixed point (or a stable spiral) which will be the state of the universe at late times, restricts the ranges of $\lambda$ and $\beta$ for each model. To determine whether these models are viable models for our universe, we calculate the cosmological parameters, the relative energy densities for matter, radiation, scalar field, curvature, the effective EOS for DM and the deceleration parameter. Here we have concentrated on analyzing whether these parameters are qualitatively similar to observations at late times (i.e. at the stable fixed point). While that may be a good way to rule out models, more quantitative analysis i.e. matching with actual values from observations are needed to claim that a particular model is viable. That is an avenue for future work. However, as we show, even this simple analysis was able to rule out certain ranges of the values of the parameters $\lambda$ and $\beta$ because they give predictions which are inconsistent with our current universe.

Having curvature, radiation and DM-DE interaction makes the phase space of our models much richer than other models discussed in literature which do not have one or more of these ingredients. All these new fixed points are saddle points. The price we pay is that the analysis of the fixed point and their stability becomes much more complicated, particularly for Model $2$. We have done the analysis of the more complicated ones using contour plots (see for eg Figs.  (\ref{figure1},\ref{figure2},\ref{figure3},\ref{figureA1})). In our analysis, we have found conditions on $\lambda$ and $\beta$ such that all of these points exist. If there are theoretical or observational reasons for excluding these one or all of these fixed points from the evolution history, a suitable parameter space can be obtained using similar graphical techniques. This will be attempted in a later work. To reiterate, we have explored one possible evolution scenario for our models. The rich phase space structure offered by our models allows for other scenarios which can be examined in detail using restrictions imposed by observations from various channels like CMB, large scale structures and DES. For the present work, our choice of parameter space is restricted only by demanding the existence of one stable and at least one unstable fixed point. 

Dynamical systems analysis allows for two different scenarios for Model $1$ in $k=-1$ cosmology. For both the scenarios we observe that the matter domination phase is missing. However features of a matter dominated saddle point may possibly be recovered from one of the new saddle points for suitable choices of $\lambda$ and $\beta$. That is one of the directions we will be exploring in the future. We get the effective EOS to be negative for both Scenario $1$ and $2$ but $q$ is negative for former and $0$ for the latter. Hence the parameter space spanned by Scenario $2$ can be discarded as a model of our universe without further analysis. This shows the power of dynamical systems analysis alluded to in the beginning. For the $k=0$ case of this model, there is only one possible scenario where the stable fixed point has negative effective EOS but to ensure that $q<0$ at late times the range of $\lambda$ has to be restricted further. 

The analysis of Model $2$ was more complicated. This model also has several new saddle points and has a matter domination phase missing from Model $1$. However, as shown in Fig. (\ref{figuree2}), larger values of $\beta$ and $\lambda$ lead to a non-accelerating universe with a negative EOS at late times while smaller values of $\beta$ and $\lambda$ lead to viable universe models with negative $q$ and $\omega_{eff}$. This is true for both open and flat models. Again, a suitable range of parameter space satisfying observational bounds will be attempted in a future work. 

One interesting feature observed in both these models is that, in the presence of DM-DE interaction,  even a dust like DM can reproduce behaviour of complicated DM models. In  Model $1$, it behaves like exotic DM at early times, dust at intermediate phase and ordinary matter at late times which is shown in Fig. (\ref{figuree1},\ref{figureevolz}). 
In Model $2$, DM behaves like stiff matter at early times, then it acts like dust at intermediate phase and exotic DM at late times which is shown in the evolution plot given in Fig. (\ref{figuree2}) for a generic value of $\lambda $ and $\beta$. The EOS of DM has a smooth variation in Model $1$ but not in Model $2$, where at certain points it becomes ill defined. 

This feature of interacting DM-DE models reproducing effects of complicated DM models needs to explored more quantitatively. This is because the effects on structure formation may be different for these although the effects look same in the context of cosmological evolution. This feature can also be explored for other choices of potentials, for other types of interactions and for other types of DM content. It would be intriguing to explore the same models for positive curvature. Alternative models like time-dependent interaction and non-minimal coupling can also provide interesting alternate models of our universe.  \\
 
\noindent {\bf{Acknowledgment:}}
P K Das would like to gratefully acknowledge the IUCAA Associateship. 
\newpage
\appendix
\section{Appendix 1}
\label{appendix1}
For the interaction form for Model 1 where $Q_I= \sqrt{\frac{2}{3}} \kappa \beta \Dot{\phi} \rho_m$, we give the explicit form of the eigenvalues in Table (\ref{table:9}).
\begin{table}[H]
\addtolength{\tabcolsep}{-5.5pt}
\renewcommand{\arraystretch}{2.1}
\centering
\begin{tabular}{| c | c | c | c | c | c | c | c | c | c | c | c |} 
 \hline
Point & Eigenvalues & Existence & Stability \\
\hline\hline
$P^+_k$ & $\left( 2, 1, 3-\lambda \sqrt{\frac{3}{2}}, 3+2\beta \right)$ &  & $\beta \le -\frac{3}{2}$ : Saddle $\forall ~ \lambda$  \\ [0.5ex]

&  & $\forall \lambda$ & $\beta > -\frac{3}{2}$ \multirow{2}{*}{\(\left\{\begin{array}{c} \\ \\ \end{array}\right.\)} Saddle point for $\lambda \ge \sqrt{6}$  \\
& & & ~ ~ ~ ~ ~ ~ ~ ~ Fully unstable for $\lambda<\sqrt{6}$  \\

$P^-_k$ & $\left( 2, 1, 3+\lambda \sqrt{\frac{3}{2}}, 3-2\beta \right)$ & $\forall \lambda$ &  $\beta < 0$ \multirow{2}{*}{\(\left\{\begin{array}{c} \\ \\ \end{array}\right.\)} Saddle point for $\lambda \ge -\sqrt{6}$ \\
& & & ~ ~ ~ ~ ~ ~ ~ ~ Fully unstable for $\lambda<-\sqrt{6}$  \\  
\hline
$P_r$ & $\left( 2, -1, 1, 1 \right) $ & $\forall \lambda$ & Saddle  \\ [0.5ex]
\hline
$P_c$ & $ \left( -2, -1, -1, 1 \right)$ & $\forall \lambda$ & Saddle \\ [0.5ex]
\hline
$P_{r\phi}$ & $\left(1, -\frac{1}{2}-\frac{\sqrt{64 -15 \lambda^2}}{2 \lambda}, -\frac{1}{2}+\frac{\sqrt{64 -15 \lambda^2}}{2 \lambda}, 1+\frac{4\beta}{\lambda} \sqrt{\frac{2}{3}} \right)$ & $\lambda>2$ & Unstable Spiral  \\ [0.5ex]
\hline
$P_{\phi}$ & $\left(\frac{\lambda^2-6}{2}, \frac{\lambda^2-4}{2}, \frac{\lambda^2-2}{2}, \frac{1}{3} (3 \lambda^2+\sqrt{6} \beta \lambda-9) \right)$ & $\lambda<\sqrt{6}$  & Stable for $\lambda \le \sqrt{2}$ \\ [0.5ex]

& & & Saddle for $\lambda > \sqrt{2}$ \\
\hline
$P_{mixed}$ & $\left(1, \frac{\sqrt{6} \lambda+8\beta}{4\beta}, \frac{-\beta-\sqrt{3}\sqrt{1-\beta^2}}{2\beta}, \frac{-\beta+\sqrt{3}\sqrt{1-\beta^2}}{2\beta} \right)$ & $\forall \lambda$, $\beta^2 > \frac{3}{4}$ & Saddle  \\ [0.5ex]
\hline
$P$ & $\left( -\frac{\sqrt{6}\lambda -4 \beta}{4 \beta}, -1, \frac{-2\beta-\sqrt{6}\sqrt{1+2\beta^2}}{2\beta}, \frac{-2\beta+\sqrt{6}\sqrt{1+2\beta^2}}{2\beta} \right)$ & $\forall \lambda$
 & Saddle \\ [0.5ex]
\hline
$P_{{m\phi}_1}$ & $\left( \frac{4\beta^2-9}{6}, \frac{4\beta^2-3}{6}, \frac{4\beta^2+3}{6}, \frac{4\beta^2+2\sqrt{6} \lambda \beta+9}{6} \right)$ & $\forall \lambda$ & Saddle \\ [0.5ex]
\hline
$P_{{m\phi}_2}$ & $\left( \frac{\sqrt{6} \lambda-4\beta}{2\sqrt{6} \lambda +4 \beta}, -\frac{\sqrt{6} \lambda+8\beta}{2\sqrt{6} \lambda +4 \beta}, \lambda_+, \lambda_- \right)$ & $\forall \lambda$ & Saddle \\ [0.5ex]
\hline

$P_{c\phi}$ & $\left( -1, -1-\frac{\sqrt{8-3\lambda^2}}{\lambda}, -1+\frac{\sqrt{8-3\lambda^2}}{\lambda}, -1+\frac{2\beta}{\lambda} \sqrt{\frac{2}{3}} \right)$ & $\lambda>\sqrt{2}$ & Stable Spiral  \\ [0.5ex]

\hline
\end{tabular}
\caption{Eigenvalues and stability of the fixed points in the parameter space of $\beta$ and $\lambda$ of Model 1.}
\label{table:9}
\end{table}

\begin{figure}[H]
    \begin{minipage}[t]{0.38\textwidth}
        \centering
        \includegraphics[width=\textwidth]{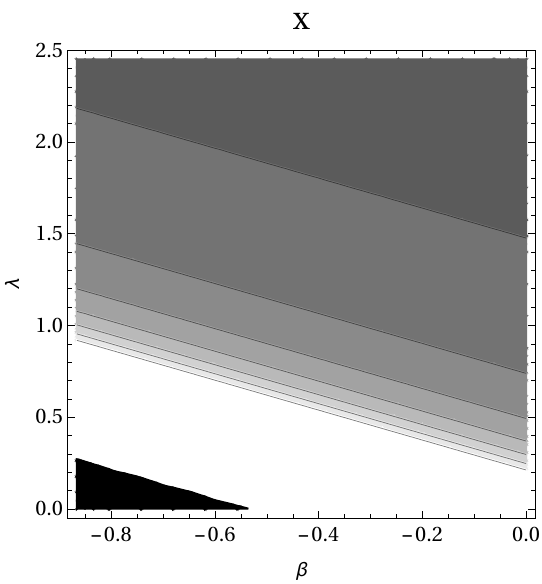} 
    \end{minipage}
    \hspace{1.7cm}
    \begin{minipage}[t]{0.38\textwidth}
        \centering
        \includegraphics[width=\textwidth]{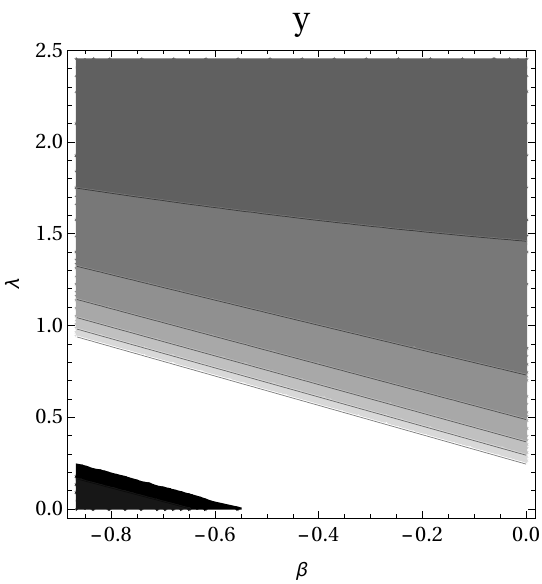}
    \end{minipage}

    \vspace{0.1cm}

    \begin{minipage}[t]{0.45\textwidth}
        \centering
        \includegraphics[width=\textwidth]{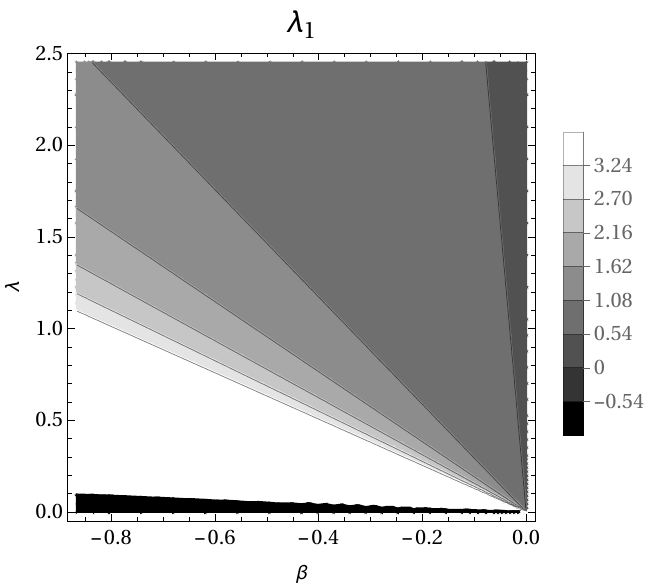} 
    \end{minipage}
    \hspace{0.5cm}
    \begin{minipage}[t]{0.45\textwidth}
        \centering
        \includegraphics[width=\textwidth]{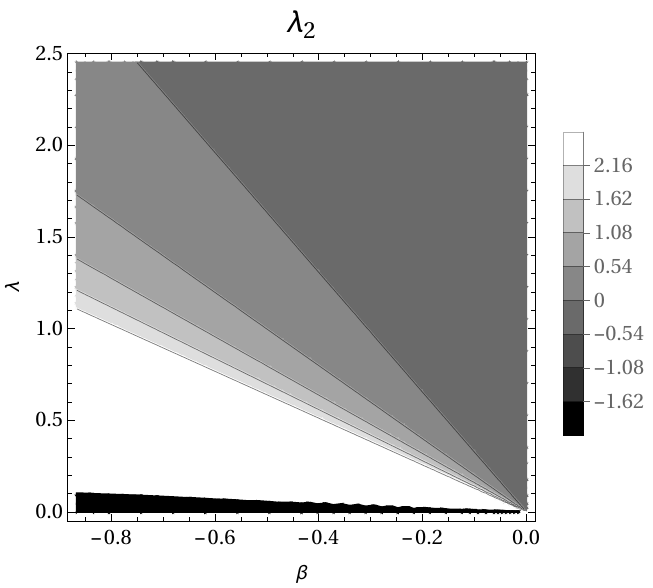} 
    \end{minipage}

    \vspace{0.1cm}

    \begin{minipage}[t]{0.45\textwidth}
        \centering
        \includegraphics[width=\textwidth]{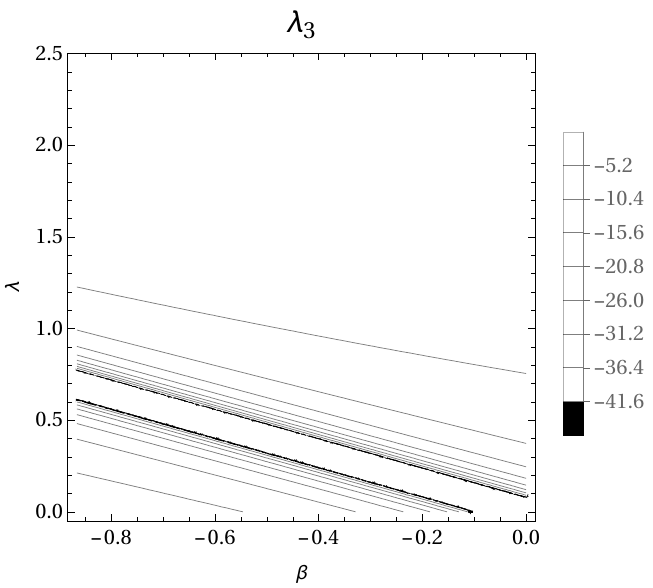} 
    \end{minipage}
    \hspace{0.5cm}
    \begin{minipage}[t]{0.45\textwidth}
        \centering
        \includegraphics[width=\textwidth]{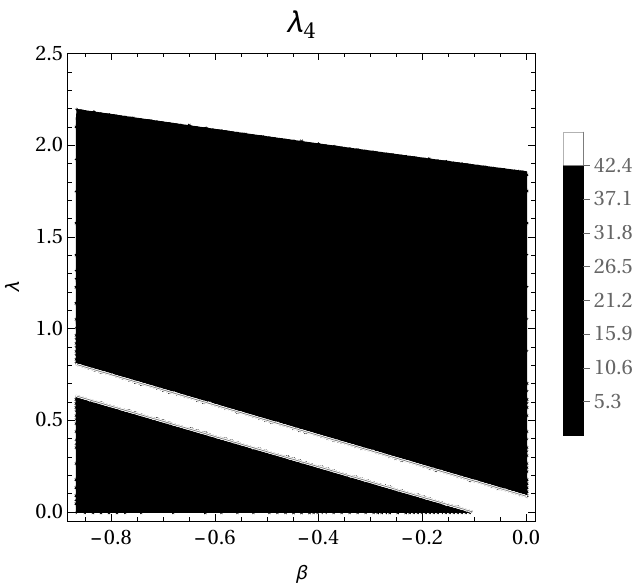} 
    \end{minipage}
    
    \caption{Plot of the dynamical variables and eigenvalues for the fixed point $P_{{m\phi}_2}$ having interaction form $Q_I= \sqrt{\frac{2}{3}} \kappa \beta \Dot{\phi} \rho_m$.}
    
    \label{figureA1}
\end{figure}

\begin{figure}[H]
    \begin{minipage}[t]{0.45\textwidth}
        \centering
        \includegraphics[width=\textwidth]{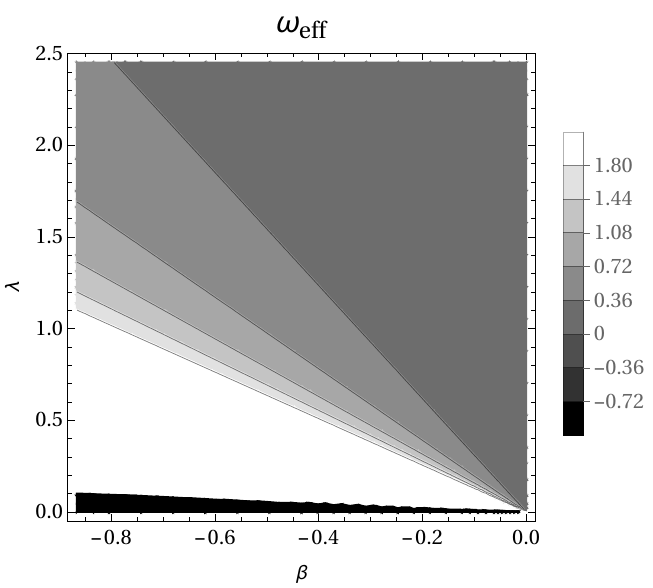} 
    \end{minipage}
    \hspace{1cm}
    \begin{minipage}[t]{0.45\textwidth}
        \centering
        \includegraphics[width=\textwidth]{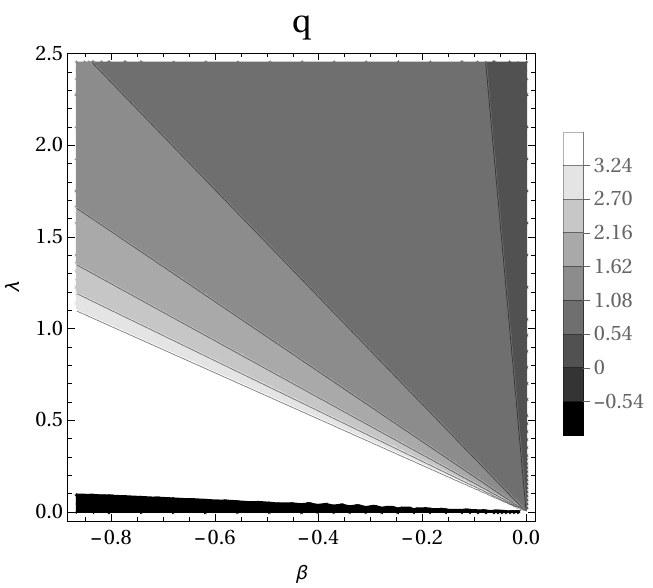} 
    \end{minipage}

    \vspace{0.1cm}
    
    \begin{minipage}[t]{0.45\textwidth}
        \centering
        \includegraphics[width=\textwidth]{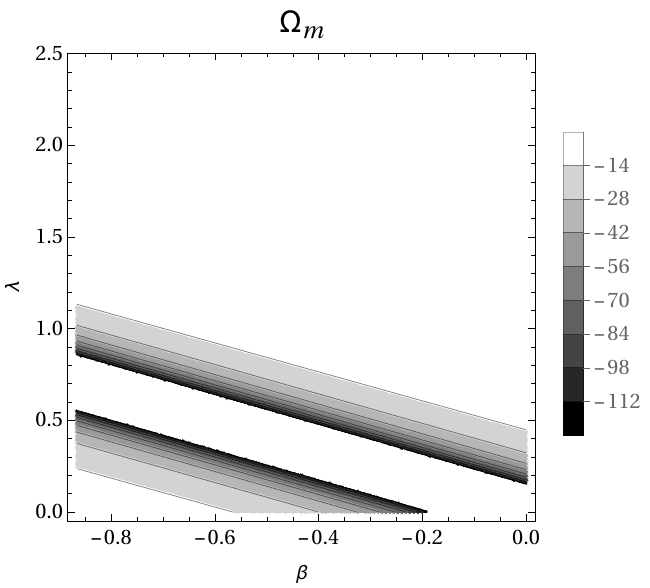} 
    \end{minipage}
    \hspace{1cm}
    \begin{minipage}[t]{0.45\textwidth}
        \centering
        \includegraphics[width=\textwidth]{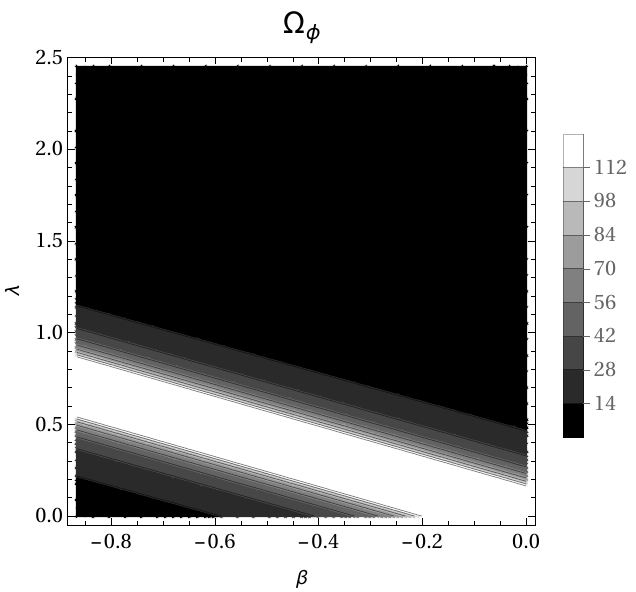} 
    \end{minipage}
    \label{figureA3}
    \caption{Plot of cosmological parameters for the interaction form $Q_I= \sqrt{\frac{2}{3}} \kappa \beta \Dot{\phi} \rho_m$.}
\end{figure}
\noindent From Table (\ref{table:1}), it is evident that for the fixed point $P_{{m\phi}_2}$, the dynamical variables are expressed in terms of $\beta$ and $\lambda$. Since it is difficult to visualise the behaviour of this point, we check it graphically. Here, we analyze the features of this fixed point following the same process we carried out for Model II. First, we check the overlap region for the dynamical variables $x$ and $y$ shown in Fig. (9). Within this overlap region, we examine the behaviour of eigenvalues presented in Fig. (10). 
The eigenvalues $\lambda_+$ and $\lambda_-$ expressions are huge and complicated to put in Table (\ref{table:9}) which is given below: 
\begin{equation*}
    \lambda_\pm = \frac{1}{2 \left(2 \beta +\sqrt{6} \lambda\right)^2}\left(-12 \beta ^2 -9 \sqrt{6} \beta  \lambda -9 \lambda^2- \Delta \right)
\end{equation*}
where
\[
\Delta = \sqrt{
\begin{aligned}
   & 720 \beta ^4+1296 \beta ^2-144 \sqrt{6} \beta  \lambda^5-1152 \beta ^2 \lambda^4-567 \lambda^4-576 \sqrt{6} \beta ^3 \lambda^3-54 \sqrt{6} \beta  \lambda^3 \\
   & - 768 \beta ^4 \lambda^2+1998 \beta ^2 \lambda^2 +1944 \lambda^2-64 \sqrt{6} \beta ^5 \lambda+936 \sqrt{6} \beta ^3 \lambda+1296 \sqrt{6} \beta  \lambda 
\end{aligned}
}
\]
As seen from Fig. (10), the eigenvalues ($\lambda_1$, $\lambda_2$, $\lambda_3$, $\lambda_4$) are both positive and negative in the overlap region which indicates that it is a saddle point.
Then, we plot the cosmological parameters corresponding to the fixed point $P_{{m\phi}_2}$. We observe that the $\omega_{eff}$, $q$, $\Omega_m$ and $\Omega_\phi$ can be both positive and negative for the overlap region of $x$ and $y$.\\

\section{Appendix 2}
\label{appendix2}
\begin{figure}[H]
    \begin{minipage}[t]{0.45\textwidth}
        \centering
        \includegraphics[width=\textwidth]{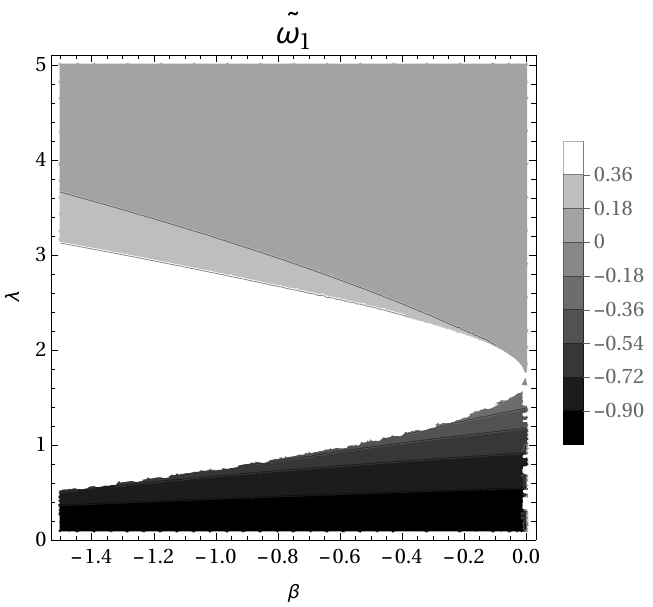} 
    \end{minipage}
    \hspace{0.5cm}
    \begin{minipage}[t]{0.45\textwidth}
        \centering
        \includegraphics[width=\textwidth]{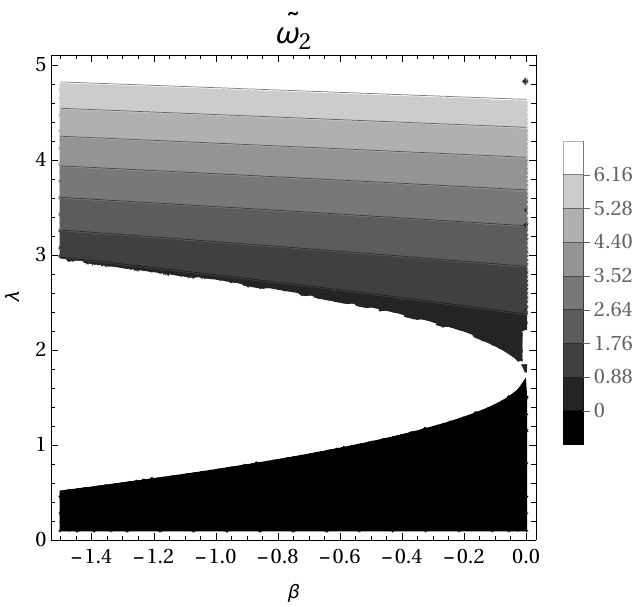}
    \end{minipage}
\caption{Plot of $\omega_{m_2}$ for the fixed points $P_{{m\phi}_2}$ and $P_{{m\phi}_3}$.}
\label{figureeosm2}
\end{figure}

In this appendix we describe the behaviour of $\omega_{m_2}$ for Model 2 at the fixed points $P_{{m\phi}_2}$ and $P_{{m\phi}_3}$. In Fig. (\ref{figureeosm2}), the contour plot of $\omega_{m_2}$ as function of $\lambda$ and $\beta$ are given for these two points. 
The region of interest is where the $x$ and $y$ coordinates of these points exist (See Fig.  (\ref{figure3}) and (\ref{figure4}) respectively). We can see that, in that region $\omega_{m_2} < 0$ for $P_{{m\phi}_2}$  and is positive for $P_{{m\phi}_3}$.

\bibliographystyle{ieeetr}
\bibliography{References}

\end{document}